\documentclass[useAMS,natbib,multicolumn,subfigure]{mn2e}
\usepackage{amsmath}
\usepackage[T1]{fontenc}
\usepackage{ae,aecompl}

\usepackage[latin1]{inputenc}  
\usepackage{graphicx,amsmath,xcolor}
\usepackage{amssymb}
\usepackage{indentfirst}
\usepackage{hyperref}
\usepackage{graphicx,pst-all}
\usepackage{tikz}
\usepackage{layout}
\usepackage{fancyhdr}

\newcommand{\kms}{\hbox{\kern 0.20em km\kern 0.20em s$^{-1}$}}
\newcommand{\cmt}{\hbox{\kern 0.20em cm$^{-3}$}}
\newcommand{\cmd}{\hbox{\kern 0.20em cm$^{-2}$}}

\def\msol{\hbox{\kern 0.20em $M_\odot$}}
\def\lsol{\hbox{\kern 0.20em $L_\odot$}}
\def\rsol{\hbox{\kern 0.20em $R_\odot$}}
\def\sr{\hbox{\kern 0.20em sr}}
\def\srmu{\hbox{\kern 0.20em sr$^{-1}$}}

\def\g{\hbox{\kern 0.20em g}}
\def\gmu{\hbox{\kern 0.20em g$^{-1}$}}
\def\kg{\hbox{\kern 0.20em kg}}
\def\pc{\hbox{\kern 0.20em pc}}

\def\mum{\hbox{\kern 0.20em $\mu$m}}
\def\mumd{\hbox{\kern 0.20em $\mu$m$^{-2}$}}
\def\cm{\hbox{\kern 0.20em cm}}
\def\m{\hbox{\kern 0.20em m}}
\def\km{\hbox{\kern 0.20em km}}
\def\nm{\hbox{\kern 0.20em nm}}

\def\s{\hbox{\kern 0.20em s}}
\def\h{\hbox{\kern 0.20em h}}
\def\sec{\hbox{\kern 0.20em sec}}
\def\min{\hbox {\kern 0.20em min}}
\def\smu{\hbox{\kern 0.20em s$^{-1}$}}
\def\smd{\hbox{\kern 0.20em s$^{-2}$}}
\def\an{\hbox{\kern 0.20em an}}
\def\anmu{\hbox{\kern 0.20em an$^{-1}$}}
\def\deg{\hbox{\kern 0.20em $^{\rm o}$}}
\def\yr{\hbox{\kern 0.20em yr}}
\def\yrmu{\hbox{\kern 0.20em yr$^{-1}$}}
\def\Myr{\hbox{\kern 0.20em Myr}}
\def\Mymu{\hbox{\kern 0.20em Myr$^{-1}$}}
\def\K{\hbox{\kern 0.20em K}}
\def\pcmu{\hbox{\kern 0.20em pc$^{-1}$}}
\def\pcmd{\hbox{\kern 0.20em pc$^{-2}$}}
\def\pcmt{\hbox{\kern 0.20em pc$^{-3}$}}
\def\kms{\hbox{\kern 0.20em km\kern 0.20em s$^{-1}$}}
\def\kmpd{\hbox{\kern 0.20em km$^{2}$}}
\def\kpc{\hbox{\kern 0.20em kpc}}
\def\cms{\hbox{\kern 0.20em cm\kern 0.20em s$^{-1}$}}
\def\erg{\hbox{\kern 0.20em erg}}
\def\ergs{\hbox{\kern 0.20em erg}}
\def\cmpd{\hbox{\kern 0.20em cm$^2$}}
\def\cmmd{\hbox{\kern 0.20em cm$^{-2}$}}
\def\cmms{\hbox{\kern 0.20em cm$^{-6}$}}
\def\cmpt{\hbox{\kern 0.20em cm$^3$}}
\def\cmmt{\hbox{\kern 0.20em cm$^{-3}$}}
\def\mpd{\hbox{\kern 0.20em m$^2$}}
\def\mmd{\hbox{\kern 0.20em m$^{-2}$}}
\def\mpt{\hbox{\kern 0.20em m$^3$}}
\def\mmt{\hbox{\kern 0.20em m$^{-3}$}}
\def\mujy{\hbox{\kern 0.20em $\mu$Jy}}
\def\mjy{\hbox{\kern 0.20em mJy}}
\def\Mj{\hbox{\kern 0.20em MJy}}
\def\jy{\hbox{\kern 0.20em Jy}}
\def\ghz{\hbox{\kern 0.20em GHz}}
\def\srmd{\hbox{\kern 0.20em sr$^{-1}$}}

\def \mum{$\mu$m}
\def\G{\hbox{\kern 0.20em G}}

\def\htwo{\hbox{H${}_2$}}
\def\h13cop{\hbox{H$^{13}$CO$^{+}$}}

\def\h2o{\hbox{H$_2$O}}

\begin{document}

\title[Astrochemical evolution with ASAI]{Astrochemical evolution along star formation: Overview of the IRAM Large Program ASAI}

\author[B. Lefloch et al.]{
Bertrand Lefloch$^{1,10}$\thanks{E-mail: bertrand.lefloch@univ-grenoble-alpes.fr},
R. Bachiller$^2$,
C. Ceccarelli$^{1}$,
J. Cernicharo$^{3}$,
C. Codella$^{4}$,
\newauthor
A. Fuente$^{2}$,
C. Kahane$^{1}$,
A. L\'opez-Sepulcre$^{6,1}$,
M. Tafalla$^{2}$,
C. Vastel$^{7}$,
E. Caux$^{7}$,
\newauthor
M. Gonz\'alez-Garc\'{\i}a$^{2,3}$,
E. Bianchi$^{4,5}$,
A. G\'omez-Ruiz$^{8,4}$,
J. Holdship$^{9}$,
E. Mendoza$^{10}$,
\newauthor
J. Ospina-Zamudio$^1$,
L. Podio$^{4}$,
D. Qu\'enard$^{9}$,
E. Roueff$^{11}$,
N. Sakai$^{12}$,
S. Viti$^{9}$,
\newauthor
S. Yamamoto$^{13}$,
K. Yoshida$^{13}$,
C. Favre$^4$,
T. Monfredini$^{14}$,
H.M. Quiti\'an-Lara$^{14}$,
\newauthor
N. Marcelino$^{3}$,
H.M. Boechat-Roberty$^{14}$,
S. Cabrit$^{15}$
\\
$^1$CNRS, IPAG, Univ. Grenoble Alpes, F-38000 Grenoble, France\\
$^2$IGN Observatorio Astronómico Nacional, Apartado 1143, 28800 Alcalá de Henares, Spain\\
$^3$Group of Molecular Astrophysics, ICMM, CSIC, C/Sor Juana Inés de La Cruz N3, E-28049, Madrid, Spain\\
$^4$INAF, Osservatorio Astrofisico di Arcetri, Largo Enrico Fermi 5, I-50125 Firenze, Italy\\
$^5$Università degli Studi di Firenze, Dipartimento di Fisica e Astronomia, Via G. Sansone 1, I-50019 Sesto Fiorentino, Italy\\
$^6$IRAM, 300 rue de la Piscine, 38406 Saint-Martin d' H\`eres, France\\
$^7$Universit\'e de Toulouse, UPS-OMP, IRAP, Toulouse, France\\
$^8$CONACYT-Instituto Nacional de Astrof\'{\i}sica, Optica y Electr\'onica, Luis E. Erro 1, 72840 Tonantzintla, Puebla, M\'exico\\
$^9$Department of Physics and Astronomy, UCL, Gower St., London, WC1E 6BT, UK\\
$^{10}$IAG, Universidade de S\~ao Paulo, Cidade Universit\'aria, SP 05508-090, Brazil \\
$^{11}$Sorbonne Université, Observatoire de Paris, Université PSL, CNRS, LERMA, F-92190, Meudon, France \\
$^{12}$The Institute of Physical and Chemical Research (RIKEN), Wako, Saitama 351-0198, Japan\\
$^{13}$Department of Physics, University of Tokyo, Bunkyo-ku, Tokyo 113-0033, Japan\\
$^{14}$Observatorio do Valongo, Universidade Federal do Rio de Janeiro, Rio de Janeiro, 20080-090, Brasil\\
$^{15}$Sorbonne Université, Observatoire de Paris, Université PSL, CNRS, LERMA, F-75014 Paris, France\\
}

\date{Accepted 2018 March 22. Received 2018 March 9; in original form 2017 November 4}
\pubyear{2018}

\label{firstpage}
\pagerange{\pageref{firstpage}--\pageref{lastpage}}
\maketitle

\begin{abstract}
Evidence is mounting that the small bodies of our Solar System, such as comets and asteroids, have at least partially inherited their chemical composition from the first
phases of the Solar System formation. It then appears that the molecular complexity of these small bodies is most likely  related to the earliest stages of star formation. It is therefore important to characterize and to understand how the chemical evolution changes with solar-type protostellar evolution. We present here the Large Program "Astrochemical Surveys At IRAM" (ASAI). Its goal is to carry out unbiased millimeter line surveys between 80 and 272 GHz of a sample of ten template sources, which fully cover the first stages of the formation process of solar-type stars, from prestellar cores to the late protostellar phase.

In this article, we present an overview of the surveys and results obtained from the analysis of the 3~mm band observations. The number of detected main isotopic species barely varies with the evolutionary stage and is found to be very similar to that of massive star-forming regions. The molecular content in O- and C- bearing species allows us to define  two chemical classes of envelopes, whose composition is dominated by either a) a rich content in O-rich complex organic molecules, associated with hot corino sources, or b) a rich content in hydrocarbons, typical of Warm Carbon Chain Chemistry sources. Overall, a high  chemical richness is found to be present already in the initial phases of solar-type star formation.
\end{abstract}

\begin{keywords}
physical data and processes: astrochemistry -- ISM: jets and outflows-molecules-abundances -- Stars:formation
\end{keywords}

\maketitle
\section{Introduction}
Understanding "our chemical origins", i.e. the evolution of matter
during the long process that brought it from prestellar cores, to protostars, protoplanetary disks, and
eventually to the bodies of the Solar System, is one of the most active topics in contemporary Astrophysics (Caselli \& Ceccarelli 2012).

Once the gravitational collapse is underway, the forming star is at the center of a thick envelope, from
where it accretes matter. Class 0 sources represent the first stages of the collapse (e.g. Andr\'e et al. 1993). The innermost envelope regions, with a size of a few 100 AU, known as {\em hot
corinos}, are heated by the radiation emitted by the central object and the ices are sublimated. The
molecules forming the ices are thus liberated and injected into the gas phase, where they may undergo
further reactions. They share similarities with hot cores but are not just scaled-down versions of them
(Bottinelli et al. 2007). Hot corinos are similar in size and composition to the nebula precursor of the
Solar System (see e.g. Jaber et al. 2014), so that their study can be considered as an archeological study of our Solar System.

Only a handful of hot corinos have been identified so far (IRAS16293-2422: Cazaux et al. 2003; IRAS2, IRAS4B: Bottinelli et al. 2007; IRAS4A: Bottinelli et al. 2004b, Taquet et al. 2015; HH212: Codella et al. 2016; L483: Oya et al. 2017; B335: Imai et al. 2016). Their nature as well as their molecular composition remain unclear: IRAS16293 is the only hot corino
investigated in detail until now (Caux et al. 2011; Jaber et al. 2014; J{\o}rgensen et al. 2012, 2016), and may not be representative of the whole Class 0. As a matter of fact, Sakai et al. (2008) have discovered a different type of chemically distinct Class 0 protostars, the so-called Warm Carbon Chain Chemistry (WCCC) sources, that are C-chain enriched, but -unlike hot corinos- poor in complex organic molecules (COMs). The actual composition of all these protostars, their similarities and differences, and their origin (that could be related to the infall dynamics) remain yet to be established.
The protostellar phase plays a major role in building up the molecular complexity, as highlighted by the recent detection of pre-biotic molecules, like e.g. glycolaldehyde (J{\o}rgensen et al. 2012) and formamide (Kahane et al. 2013) around the solar-type protostar IRAS~16293-2422.

Evidence is mounting that the small bodies of our Solar System, such as comets and asteroids, have  at least partially inherited their chemical composition from the first
phases of the Solar System formation. For example, the molecular abundances in comet Hale-Bopp were
found to be similar to those in the protostellar outflow L1157 (Bockel\'ee-Morvan et al. 2000) ,
and the HDO/H$_2$O ratio measured in the ice of comets is, within a factor of two, equal to the ocean value (Mumma \& Charnley, 2011; Ceccarelli et al. 2014). Moreover, the large deuteration of amino acids in meteorites suggests that at least a fraction
of them were formed during the first phases of the Solar System (Pizzarello \& Huang 2005).

Thanks to the recent spectacular progress of radioastronomical and far infrared observatories, detailed
observations of systems at different evolutionary stages have become possible, and these can shed light on the most decisive chemical processes determining the evolution. However, despite a wealth of fragmentary studies in the literature, the characterization and understanding of the
chemical evolution along solar-type protostellar evolution are far from being achieved.

Systematic spectral line surveys constitute the most powerful diagnostic tool to carry out a comprehensive study of the chemical evolution of star-forming regions. In general terms, as transitions with different
upper energy levels and Einstein coefficients are excited at different temperatures and densities, line surveys permit to efficiently probe different regions along the line of sight (see e.g. Tercero et al. 2010; Caux et al. 2011). Star-forming regions are particularly complex because of their strong spatial chemical differentiation and because different lines from the same species are excited under different conditions (temperature, density, velocity field, etc), with sometimes complex kinematics, where infall and outflow motions are simultaneously present.

Unbiased spectral surveys of low-mass, solar-type objects have been carried out so far only towards TMC1 and Barnard~1 (Marcelino et al. 2009; Cernicharo et al. 2012), IRAS16293 (Blake et al. 1994; Caux et al. 2011; J{\o}rgensen et al. 2016), IRAS4 (Blake et al. 1995), L1527 at 3~mm only (Takano et al. 2011) and R CrA IRS7B at 0.8mm (Watanabe et al. 2012). Now, the new capabilities of the IRAM 30m telescope have made it possible to take a major step forward in the investigation of molecular complexity
along with star formation, by observing with unprecedented sensitivity the emission of molecular rotational
transitions in the millimeter domain, from 80 GHz to 272 GHz, in a greatly reduced amount of time.

With all this in mind, we have undertaken a
Large Program called ASAI\footnote{http://www.oan.es/asai/} (Astrochemical Studies At IRAM) to characterize and to understand the chemical evolution along solar-type protostellar evolution. To do so, we  have used the IRAM 30m telescope to carry out unbiased millimeter line surveys of a sample of template sources, which cover the full formation process of solar-type stars, from prestellar cores to protoplanetary disks. The objective of this
paper is to describe the general characteristics of the survey: observational strategy, methods of the data analysis and features of the target, together with some overall and first results (some of them already published in dedicated papers) illustrating the richness of the obtained data.

In this presentation article of the project, we have focused on the molecular content of the 3~mm band (80--116 GHz).  It is the only band which was observed towards all the template sources, with the exception of AB Aur. The systematic approach of our Large Program allows us to draw some first conclusions on the molecular richness of sources as a function of their evolutionary status. The detailed ASAI data analysis including the bands at 2~mm and 1.3~mm is still under way. Further results on the different sources, transversal studies, as well as a full analysis of the survey will be presented in forthcoming papers.

\section{The source sample}

We have selected a sample of ten template sources illustrating the different chemical stages a solar-type star undergoes during its formation process, presented in Table~1.

{\bf TMC1} is a quiescent dense ridge in Heiles Cloud 2 at d $\simeq$ 140 pc (Cernicharo \& Gu\'elin 1987) in the Taurus molecular cloud complex. It is characterized by a very rich molecular spectrum and strong chemical differences along the ridge, and is considered as an ideal source to study the chemistry of dark clouds (e.g. Pratap et al. 1997, Liszt \& Ziurys 2012). The position observed here, 'the cyanopolyyne peak', is particularly rich in carbon chains, including radicals and cyanopolyynes, and it was previously surveyed in the 3-mm band with the IRAM 30m radiotelescope (see Marcelino et al. 2007, 2009; Cernicharo et al., 2012). Among the molecules detected in these previous surveys with the 30m IRAM telescope it is worth noting  the completely unexpected saturated propene molecule CH$_3$CHCH$_2$ (Marcelino et al., 2007).

{\bf L1544} is a starless core in the Taurus molecular cloud complex (d $\simeq$ 140 pc; Cernicharo \& Gu\'elin 1987) on the verge of the gravitational collapse (Caselli et al. 2012, and references within). It is considered the prototype of prestellar cores. Its central high density ($2\times 10^{6}\cmmt$) and very low temperature ($\sim$7 K) generate a chemistry typical of the interiors of dark clouds in which CO is depleted and the deuterium fractionation is high, although differentiated chemical processes can take place in the external layers (Caselli et al. 1999; Vastel et al. 2006, 2014, 2015b, 2016).

{\bf Barnard 1} is a dense core in the Perseus molecular complex (Bachiller \& Cernicharo 1984, 1986) at d$\simeq$ 230 pc (Hirano \& Liu, 2014). It contains several active star-forming sites of which the position studied here, B1b, is one of the most interesting because of its rich molecular spectrum (Bachiller et al. 1990a; Cernicharo et al. 2012; Daniel et al. 2013).  B1b was previously surveyed in the 3~mm band with the IRAM 30m radiotelescope by Marcelino et al. (2009) and Cernicharo et al. (2012). It consists of two objects B1-bN and B1-bS, separated by approximately $18\arcsec$  (Pezzuto et al. 2012; Gerin et al. 2015), which have been proposed as candidates for the first hydrostatic core (FHSC) stage based on the properties of their spectral energy distribution and their outflows (Gerin et al. 2015; Fuente et al. 2017). Both sources have a low luminosity of $0.28\lsol$ and $0.49\lsol$, respectively (Pezzuto et al. 2012).
The targeted position lies approximately halfway between B1-bN and B1-bS (see Table~1). We have  reported in Table~1 the total luminosity of the system ($0.77\lsol$). Several molecular species have been detected, some of them of high importance for astrochemistry, such as NH$_3$D$^+$  (Cernicharo et al., 2013), CH$_3$O (Cernicharo et al., 2012) and HCNO, the high energy isomer of isocyanic acid (Marcelino et al. 2009). The source exhibits a large deuteration, as evidenced by the detection of multiply deuterated species such as  D$_2$CS (Marcelino et al. 2005), ND$_2$H (Roueff et al. 2005) and ND$_3$ (Lis et al. 2002). Also, a very high abundance in sulfur compounds was reported by Marcelino et al. (2005) and  Fuente et al. (2016).

{\bf L1527} is a dark cloud in the Taurus molecular complex (d $\simeq$ 140 pc) containing IRAS04368+2557, a protostar with a luminosity of 2.75 L$_\odot$ (Tobin et al. 2013). Recently,  Robitaille et al. (2007) and Tobin et al. (2013)  proposed that L1527 probably lies at an intermediate stage Class 0/I, more evolved than previously thought. The position observed here is the nominal position of the protostar from which emerges a highly collimated molecular outflow (Hogerheijde et al. 1998). The complex kinematical structure of the region around the protostar has been recently investigated with ALMA by Oya et al. (2015) and Sakai et al. (2017). This source is considered as a prototypical warm-carbon-chain-chemistry (WCCC) source (Sakai et al. 2008, 2010; Sakai \& Yamamoto 2013).

{\bf NGC1333-IRAS4A} is a binary Class 0 source located in the Perseus molecular complex at d
$\simeq 260\pc$ (Schlafly et al. 2014). The two components, IRAS4A1 and IRAS4A2, have a separation of about $1.8\arcsec$ ($\sim $420 AU) and a total luminosity of $9.1\lsol$ (Karska et al. 2013). The IRAS4A system is associated with a spectacular large-scale (a few arcminutes) bipolar molecular outflow (e.g. Blake et al. 1995; Choi 2001, 2005; Yildiz et al. 2012; Santangelo et al. 2014). High-angular resolution observations by Santangelo et al. (2015) have disentangled two distinct molecular jets powered by each of the two components.  IRAS4A was the subject of a spectral line survey  between 200 and 400 GHz with the Caltech Submm Observatory (CSO) and the James Clerck Maxwell Telescope (JCMT) by Blake et al. (1995). The weakness of the lines made it impossible to obtain a complete coverage of the spectral windows.  Thanks to more recent, sensitive observations with the IRAM 30m telescope, IRAS4A was identified as a  hot corino protostar, the second one  after IRAS16293-2422 (Bottinelli et al. 2007). Santangelo et al. (2015) and  L\'opez-Sepulcre et al. (2017) confirmed that only source IRAS4A2 is a hot corino protostar,  in agreement with the previous detections of dimethyl ether (CH$_3$OCH$_3$), ethyl cyanide (C$_2$H$_5$CN) and water  (H$_2^{18}$O) by Persson et al. (2012); also, de Simone et al. (2017) reported the detection of glycolaldehyde (CH$_2$OHCHO). Interestingly, no COM emission is detected towards IRAS4A1.

{\bf L1157mm (IRAS 20386+6751)} is the Class 0 source driving the powerful L1157 molecular outflow (Bachiller et al. 1993, 2001; Gueth et al. 1996; Tafalla et al. 2015) in a quite isolated dark cloud in Cepheus at d$\simeq 250\pc$ (Looney et al. 2007). The protostar of $\sim 3\lsol$ is surrounded by a circumstellar disk and a protostellar envelope embedded in filamentary cloud (Gueth et al. 2003; Looney et al. 2007).

{\bf SVS13A} is part of the multiple system NGC1333-SVS13 (d$\simeq 260\pc$; Schlafly et al. 2014), where three millimetre sources, called A, B, and C, have been identified by interferometric observations (Bachiller et al. 1998; Looney et al. 2007). The angular distance between A and B is $15\arcsec$, while C is $20\arcsec$ away from A. The luminosity of SVS13-A has been estimated to be $34\lsol$ (Tobin et al. 2016). SVS13-A itself is a close (70AU) binary (VLA4A, VLA4B)in the  radio cm (Tobin et al 2016). Although SVS13-A is still deeply embedded in a large-scale ($\sim 6000$ AU; e.g. Lefloch et al. 1998a) envelope, its extended (> 0.07 pc) outflow, associated with the HH7-11 chain (e.g. Lefloch et al. 1998b, and references therein), and its low L$_{submm}$/L$_{bol}$ ratio ($\sim 0.8\%$) lead to the classification as a Class I source (e.g. Chen et al. 2009 and references therein).

{\bf AB Aur} is one of the best-studied Herbig Ae stars that host a prototypical Herbig Ae disk. Located at d $\simeq 145\pc$ in Taurus, the star has a spectral type A0-A1, a mass $\sim 2.4\msol$, and a $T_{eff}$  $\simeq 9500\K$ (van den Ancker et al. 1998). The disk around AB Aur shows a complex structure: it exhibits an asymmetric
dust ring at R= 70-140 AU from the star (Pi\'etu et al. 2005, Tang et al. 2012), outer spiral-arm features traced by the CO and its isotopologues emission, and inner spiral arms connecting the dusty ring with the star that have been recently imaged by Tang et al. (2017).
The chemical structure of the disk has been studied at high resolution in selected molecular lines by Schreyer et al. (2008), Fuente et al. (2010), Guilloteau et al. (2013) and Pacheco-V\'asquez et al. (2015, 2016).

{\bf L1157-B1} is a bright hot-spot in the southern blue-shifted lobe of the powerful L1157 outflow (d$\simeq 250\pc$; Looney et al. 2007). This outflow is considered the archetype of the so-called chemically active outflows (Bachiller \& P\'erez-Guti\'errez 1997, Bachiller et al. 2001, Burkhardt et al. 2016). The richness and intensity of the molecular spectrum from L1157-B1 has made this position a testbed for shock-chemistry models and a favoured target for molecular surveys (e.g. Gusdorf et al. 2008; Codella et al. 2010, 2015; Lefloch et al. 2010, 2012; Yamaguchi et al. 2012; Busquet et al. 2014; Gómez-Ruiz et al. 2015; Holdship et al. 2016).
In particular, the emission of the shock region in the 3~mm band was observed recently in an unbiased manner   with the 45m telescope of the Nobeyama Radio Observatory (NRO) by Yamaguchi et al. (2012) and Sugimura et al. (2011).

{\bf L1448-R2} is a bright spot in the southern redshifted lobe of the bipolar molecular outflow L1448 (Bachiller et al. 1990b) which is emanating from a $\sim 11\lsol$ Class 0 protostar in the Perseus molecular complex at d$\simeq 235\pc$. High angular resolution observations in SiO lines (Dutrey et al. 1997) were found in agreement with a structure of incomplete and fragmented bow-shocks caused by discrete episodes of ejection from the central star ('molecular bullets'). Strong H$_2$O and high-J CO line emission was detected toward this and other L1448 bright spots by Nisini et al. (2000, 2013), Santangelo et al. (2012) and Gomez-Ruiz et al. (2013) indicating the presence of very dense ($\sim 10^6\cmmt$)  molecular gas at temperatures in excess of $500\K$.

As shown in this short presentation of the different sources of the sample, these objects have all been the subject of previous molecular gas studies at millimeter wavelengths. Maps of the molecular emission and the velocity field obtained with both millimeter single-dish telescopes and interferometers, are available for several molecular species and were used to model the origin of their emission.

\section{Observations}
\begin{table*}
\center{
\caption[]{Source and observational characteristics. The first 4 columnes refer to the source name, coordinates, distance and luminosity. The following columns report the observed spectral windows and the range of rms noise achieved in a  $1\kms$ channel for each spectral window. The rms values (in $T_A^{*}$) are measured in the range 86--87 GHz and 113--114 GHz at 3 mm, 132-133 GHz and 169--170 GHz at 2 mm, 220-221 and 260-261 GHz at 1 mm.}
\begin{tabular}{lllllllll} \hline
Sources    &  Coordinates   &  d        &   Lum.   &  3 mm     & 2 mm  & 1.3 mm & $\delta \nu$   & Comment \\
           &     (J2000)    &   (pc)    &   ($\lsol$)& (mK)   & (mK)  & (mK)  &  (kHz)        & \\  \hline
TMC1       &  $04^h41^m41.90^s$ $+25\deg 41\arcmin 27.1\arcsec$   & 140    &    --      & --    &  4.2--4.2& --     & 48.8, 195.3 & Early prestellar core  \\
L1544       &  $05^h04^m17.21^s$ $+25\deg10\arcmin 42.8\arcsec$    &  140    &    --     &2.1--7.0   &    --   & --         &  48.8     & Evolved prestellar core  \\
B1b &  $03^h33^m20.80^s$ $+31\deg 07\arcmin 34.0\arcsec$    & 230    &  0.77 &2.5--10.6(*)  &  4.4--8.0   & 4.2--4.6    &  195.3  & First Hydrostatic Core         \\
L1527       &  $04^h39^m53.89^s$ $+26\deg 03\arcmin 11.0\arcsec$  & 140    &  2.75 & 2.1--6.7(*)    &  4.2--7.1  & 4.6--4.1    &    195.3   & Class 0/I WCCC    \\
IRAS4A   & $03^h29^m10.42^s$ $+31\deg 13\arcmin 32.2\arcsec$   & 260    &  9.1 &  2.5--3.4        & 5.0--6.1   & 4.6--3.9   &   195.3  & Class 0 Hot Corino  \\
L1157mm & $20^h39^m06.30^s$ $+68\deg 02\arcmin 15.8\arcsec$  & 250     &  3 & 3.0--4.7        &  5.0--6.5   &  3.8--3.5  &    195.3     &  Class 0     \\
SVS13A   & $03^h 29^m 03.73^s$ $+31\deg 16\arcmin 03.8\arcsec$ & 260  &34 & 2.0--4.8         &  4.2--5.1    &  4.6--4.3  &   195.3  & Class I       \\
AB Aur ($\dag$)& $04^h55^m45.84^s$ $+30\deg 33\arcmin 33.0\arcsec$ &  145 & --  & 4.6--4.3& 4.8--3.9  &  2.1--4.3   &   195.3   & protoplanetary disk   \\
L1157-B1 & $20^h39^m10.20^s$ $+68\deg 01\arcmin 10.5\arcsec$ & 250 & -- & 1.1--2.9        & 4.6--7.2    &  2.1--4.2  &   195.3   & Outflow shock spot      \\
L1448-R2 & $03^h25^m40.14^s$ $+30\deg 43\arcmin 31.0\arcsec$  & 235 & -- & 2.8--4.9        & 6.0--9.7    &  2.9--4.9   &   195.3  & Outflow shock spot      \\
\hline
\end{tabular}

(*)3 mm coverage until 112GHz only. rms is measured in the range 110-111 GHz.
($\dag$) uncomplete coverage for AB Aur: 85--96 GHz, 134--145 GHz,  200--208 GHz, 216.4--228 GHz, 232.6--240GHz, 248.3--256.2 GHz, 264.2--272 GHz
}
\end{table*}

Spectral line surveys of the full source sample were carried out with the IRAM 30m telescope over six semesters from September 2012 to March 2015, using the broad-band EMIR receivers E090, E150 and E230, connected to the Fast Fourier Transform Spectrometers (FTS) either in the high (50 kHz) or low (200 kHz) spectral resolution mode. The observations were carried out in Wobbler Switching Mode, with a throw of $3\arcmin$, in order to ensure a flat baseline across the spectral bandwidth observed. The instrumental setup was decided according to the sources:
\begin{itemize}
\item {\bf TMC1}: full coverage of the 3 mm band with a 50 kHz spectral resolution was obtained previously by another team (see e.g. Marcelino et al. 2007, 2009) and will be published elsewhere (Cernicharo {\em in prep.}) Only the 2 mm band was observed. The lower-frequency range 130-150 GHz was observed at 50 kHz spectral resolution, and the high-frequency band 150-170 GHz was observed with the FTS in its 200 kHz resolution mode.
\item {\bf L1544}: only the band 72 to 115 GHz was observed, using the  FTS in its 50 kHz resolution mode.
\item {\bf Protostars and Outflow Shocks} (B1b, L1527, IRAS4A, L1157mm, SVS13A, L1157-B1, L1448-R2): the 3~mm (80--116 GHz) and 2~mm (126--170 GHz) bands were observed simultaneously. The 1.3~mm (200-272 GHz) band was covered observing  both LSB and USB simultaneously using the FTS in its 200 kHz resolution mode. Complementary observations of the band 72--80 GHz were obtained in January 2016 for L1527, IRAS4A, L1157mm, SVS13A and L1157-B1.
\item {\bf AB Aur}: the spectral line density is so low that it was decided to perform very deep integrations of a few spectral bands of 4--8 GHz only, instead of covering the full spectral windows. Observations and results are presented in Pacheco-V\'asquez et al. (2015).
\end{itemize}

Whereas E090 and E230 are 2SB receivers, most of the observations in the 2 mm band were carried out with the "old" E150 receiver, providing an effective instantaneous bandwidth of 4 GHz in double polarization H and V. A 2SB receiver at 2 mm  was installed at the IRAM 30m telescope in Summer 2014, allowing  full coverage of the band 130--150 GHz for TMC1.
Observations were carried out with very good to excellent, stable weather conditions. The half power beamwidth of the IRAM 30m telescope (HPBW) is described with a good accuracy by the HPBW$(\arcsec)$= 2460/Frequency(GHz). The typical beamwidth of the ASAI observations lies therefore in the range $21\arcsec$ -- $34\arcsec$, $14\arcsec$ -- $19\arcsec$,
$9\arcsec$ -- $12\arcsec$ at 3mm, 2mm and 1.3mm, respectively.

Pointing was carefully checked every hour on bright nearby continuum sources: NGC7538 (L1157-B1, L1157-mm), 0316+413 and 0333+321 (B1, IRAS4A, SVS13A), 0439+360 (TMC1, L1527, L1448-R2). Pointing corrections were always less than $3\arcsec$. Focus corrections were done at the beginning of each observing session. They were repeated every 6 hours, or when relevant temperature changes occurred (e.g. after sunrise or sunset). Focus corrections were always below 1~mm.

Image band rejection of the EMIR receivers usually ranges between -10 and -20 dB, depending on the frequency (see e.g. IRAM report by D. Maier, 2014).  The sensitivity of the ASAI observations is so high that line contamination from the image band cannot be ignored.  In order to permit easy identification of lines from the image band in the spectrum, it was decided to regularly shift the frequency of the Local Oscillator (LO) by a fixed amount of 50 MHz. Spurious features in the spectral bands were identified and subsequently corrected for, in close collaboration with the IRAM receiver engineers group (see e.g. IRAM report by Kramer et al. 2014 for more details).

Contamination by the nearby reference position was observed towards IRAS4A, SVS13A, L1448-R2 and L1527 for  bright molecular lines like CO, $^{13}$CO, HCO$^{+}$, HCN, for instance. The corresponding spectral settings were re-observed in position switching mode with a "clean" reference position, checked to be free of emission.

Data reduction was performed using the GILDAS/CLASS software\footnote{http://www.iram.fr/IRAMFR/GILDAS/}. A simple flat baseline was first subtracted to each spectra and spectra with very high noise were
discarded. From comparing every scan with their (50 MHz LO) frequency shifted counterpart, spurious signals from the image band could be identified and removed. The resulting scans were then averaged in order to produce the final spectral bands.
In the three millimeter bands, the final rms lies in the range 2mK and 5mK per interval of $1\kms$, depending on the frequency.

For each source, the observed position, the nominal spectral resolution of the observations, and the rms achieved in the covered spectral bands are summarized in Table~1. The official ASAI repositories are
at OAN\footnote{http://www.oan.es/asai} and IRAM\footnote{http://www.iram.fr/ILPA/LP007/}. The ASAI database provides fully reduced, calibrated spectra.

The achieved rms of the ASAI data is similar to that of the TIMASS survey of IRAS16293-2422 (Caux et al. 2011). As mentionned in Sect.~2, two sources of our sample have been previously surveyed in the same spectral range: IRAS4A and L1157-B1. In the first case, the JCMT-CSO survey by Blake et al. (1995) yielded a typical rms of about 100 mK. In the second case, the NRO 45m survey by Yamaguchi (2012) and Sugimura et al. (2011), yielded an rms of 4--10mK per channel of 1 MHz. By comparison, the rms of the ASAI is $\approx 1-3$mK per element of resolution of 0.3 MHz.

\begin{figure*}
\begin{center}
\includegraphics[width=1.95\columnwidth]{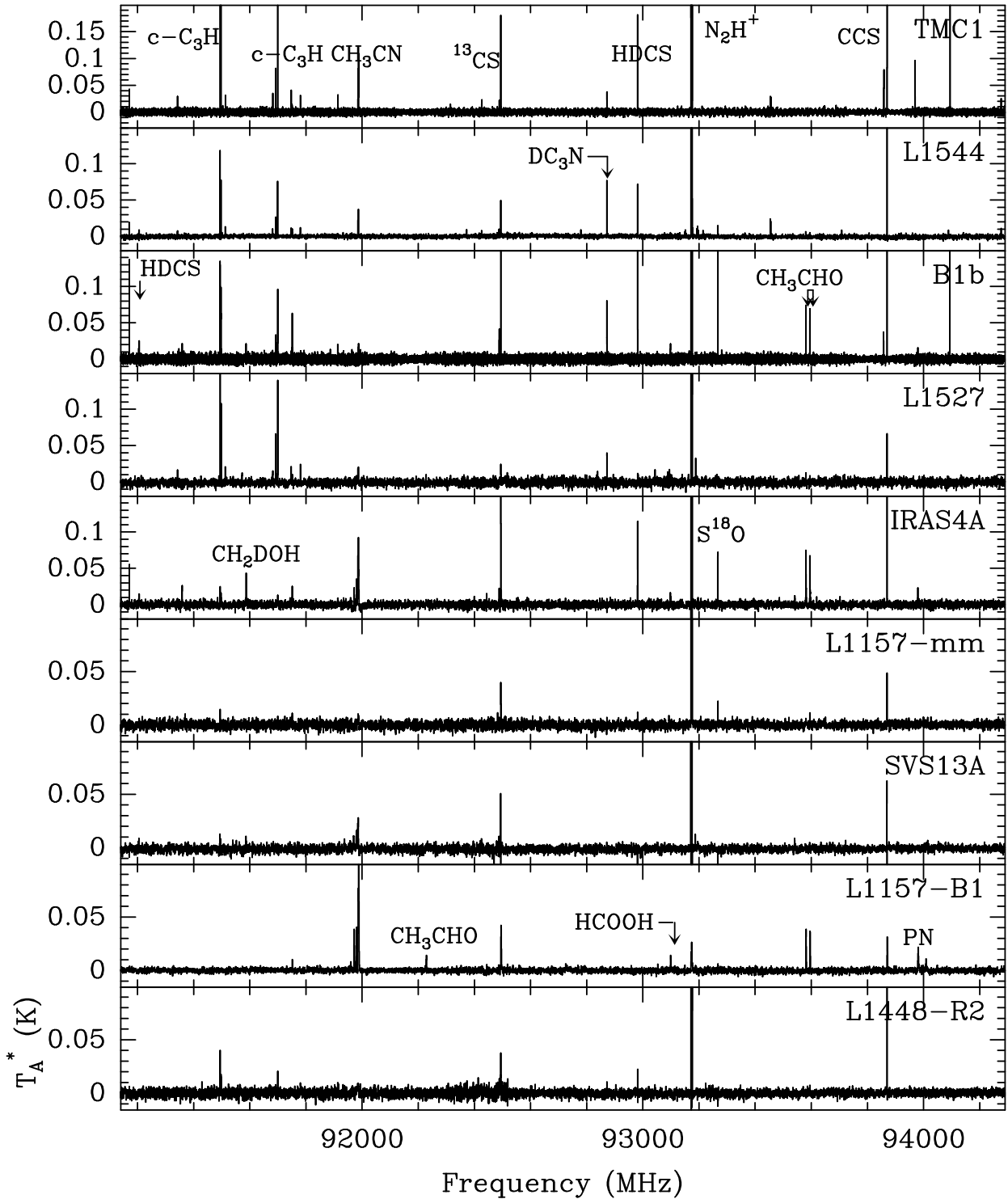}
\caption[]{Molecular line emission detected with ASAI in the spectral bands: 91100--94300 MHz. The spectral resolution is 350kHz for all sources, except for TMC1 (100 kHz). The TMC1 data were taken from Marcelino et al. (2009).  }
\end{center}
\end{figure*}

\begin{figure*}
\begin{center}
\includegraphics[width=1.95\columnwidth]{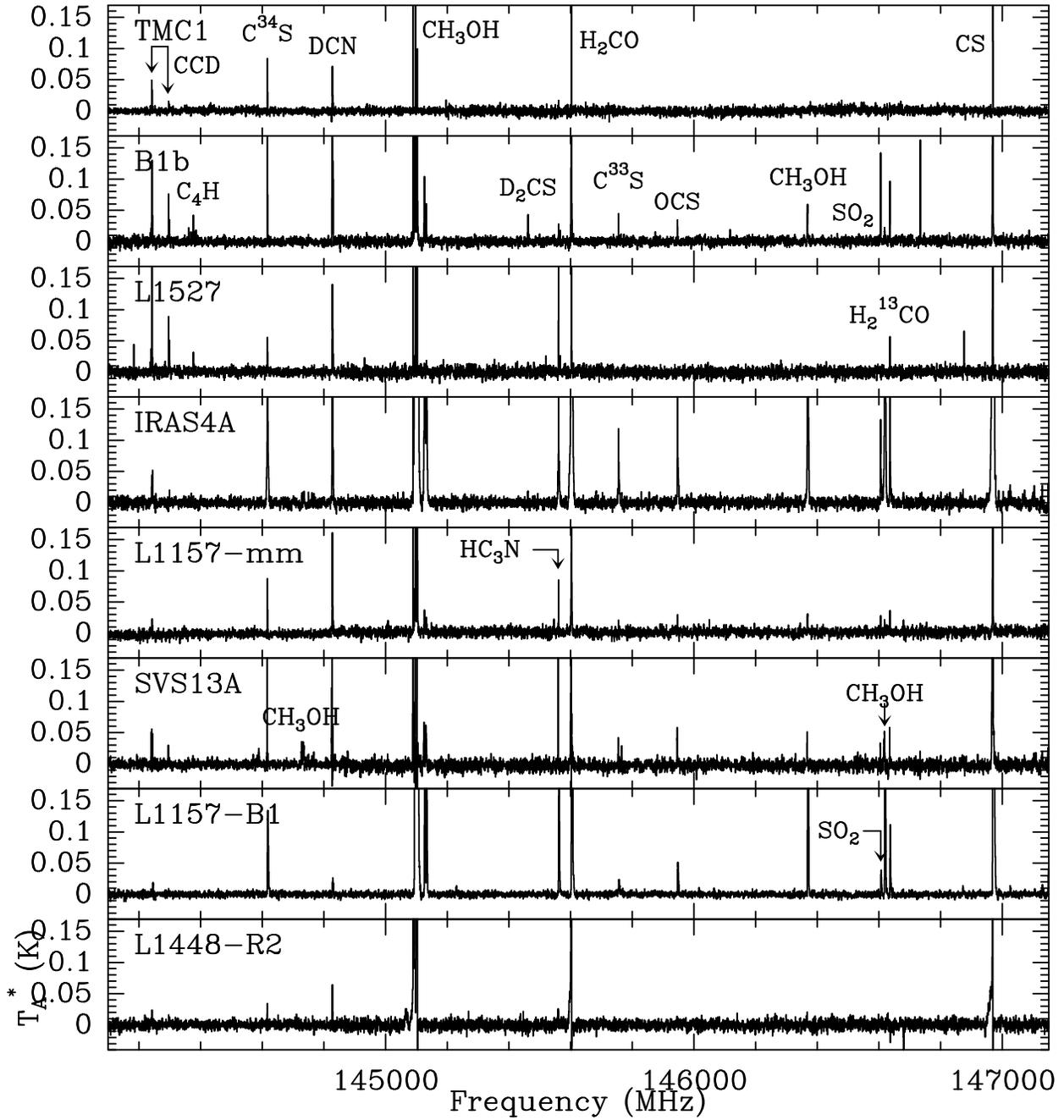}
\caption[]{Molecular line emission detected with ASAI in the spectral bands 144100--147200 MHz. The spectral resolution is 500 kHz for all the sources. }
\end{center}
\end{figure*}

\begin{figure*}
\begin{center}
\includegraphics[width=1.95\columnwidth]{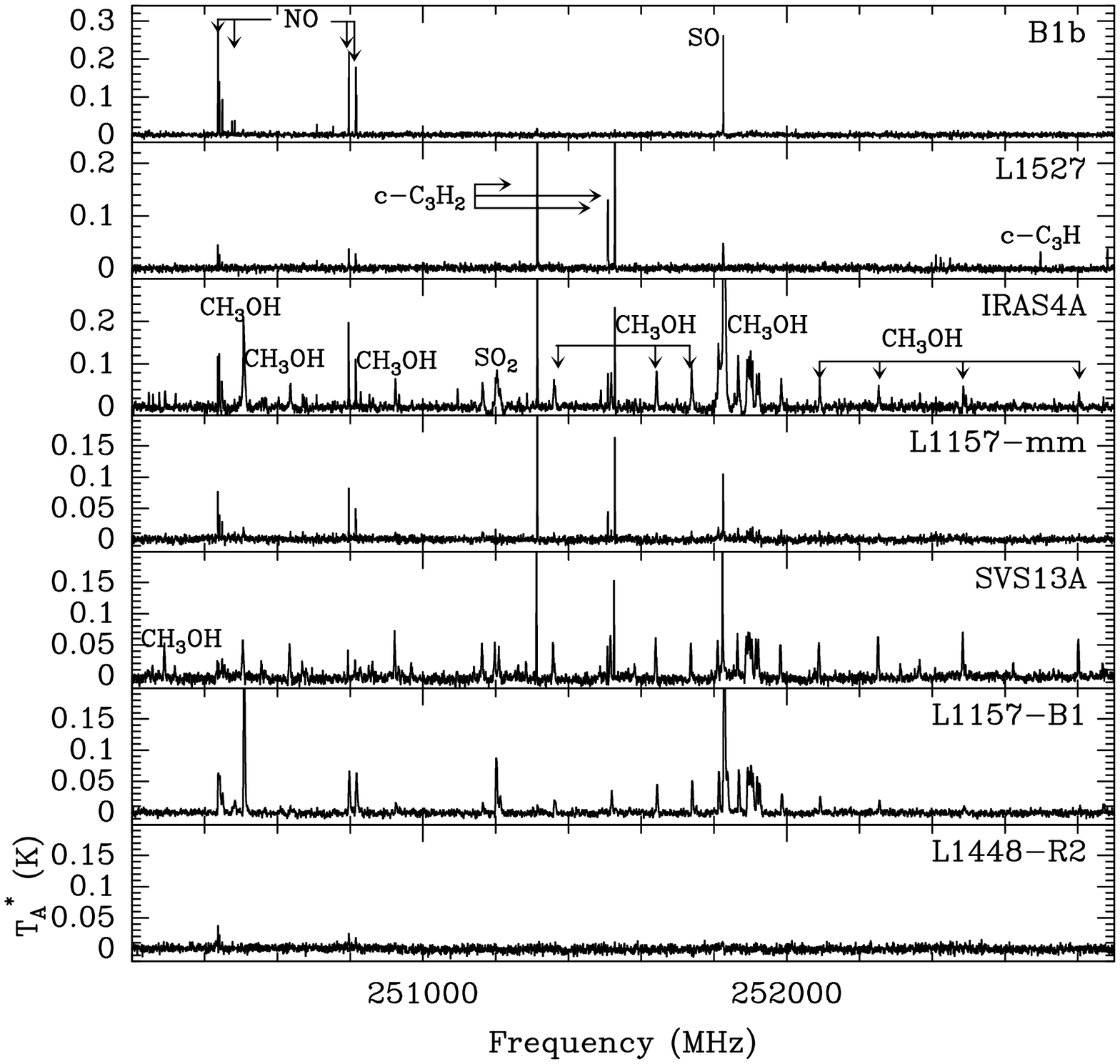}
\caption[]{Molecular line emission detected with ASAI in the spectral bands 250200--252900 MHz. The spectral resolution is 700 kHz for all the sources, except L1157-B1 (780 kHz). }
\end{center}
\end{figure*}

\section{Overall results}


The spectra of the different targets exhibit very significant differences, reflecting the evolutionary stage and the physical conditions of the different sources. This is illustrated in Figs.1--3, which display the emission in the spectral bands 91.1--94.5 GHz (Fig.~1), 144.1--147.2 GHz (Fig.~2), 250.2--252.9 GHz (Fig.~3). An example is provided by  the emission of the cyano methyl radical (CH$_2$CN) lines at 100600 MHz (se Fig.~A1). Bright CH$_2$CN emission is detected mainly towards the early sources (prestellar cores TMC1 and L1544, first hydrostatic core candidate B1b) and the WCCC protostar L1527. There is a striking anticorrelation with the emission of methyl cyanide CH$_3$CN, a chemically related species, whose emission is extremely bright towards the more evolved and luminous sources (IRAS4A, SVS13A, L1157-B1) and faint towards the early sources.  Only the low-excitation transitions of CH$_3$CN are detected towards the CH$_2$CN-rich sources, and their lines are much narrower than those of CH$_3$CN. A detailed modelling of the CH$_2$CN and CH$_3$CN emission in the ASAI sample is currently going on (Vastel in preparation).

Comparison of the COM emission in the range 90100--90258 MHz provides another interesting example of the chemical differentiation which can be observed between the sources of the sample (Fig.~A2). This band was chosen as it contains several rotational transitions of ethanol (C$_2$H$_5$OH), methyl formate CH$_3$OCHO, and  formic acid (HCOOH). The signatures of COMs are absent of the PSC (TMC1 and L1544) and the WCCC sources (L1527 and L1157-mm), and SVS13A (Class I). On the contrary, they are detected with a very good SNR towards B1b, IRAS4A and L1157-B1. Linewidths increase markedly from FHSC to Class 0 and to the shock region L1157-B1.
It is remarkable that such bright COM emission is detected towards the outflow shock region L1157-B1 whereas no emission is detected in the envelope of the protostar L1157-mm and the initial gas and dust chemical conditions are very similar. The diversity of spectral signatures was also found to reflect the diversity of excitation and chemical conditions among the sample sources. This is well illustrated by the analysis of the CO line profiles detected towards the shock region L1157-B1 by Lefloch et al. (2012).

We used the CASSIS\footnote{http://cassis.irap.omp.eu} software (Vastel et al. 2015a) for the line identification, using the CDMS\footnote{http://www.astro.uni-koeln.de} (M\"uller et al. 2005) and JPL\footnote{http://spec.jpl.nasa.gov} databases (Pickett et al. 1998). We considered
all the lines detected with an intensity higher than $3\sigma$.

We examine here how much the molecular content differs between sources, based on the molecular line identification in the 3~mm band.

\subsection{Molecular content}

\begin{table*}
\centering{
\caption[]{3~mm band observations: Census of detected molecular species classified per elemental family: oxygen, nitrogen, carbon, sulfur. In a few cases marked with $\dag$, species were detected through the emission of their rare isotopologues (D, $^{13}$C).
O- and N-bearing COMs are marked in boldface. For each molecular species, we indicate the associated TAG, according to the database used for identification (CDMS or JPL). In the last column, we have reported the molecular content of the IRAS16293-2422 envelope based on the TIMASS survey (Caux et al. 2011). }
\begin{tabular}{llcccccccccc}
\hline
  &        &  L1544 & B1b       & L1527& IRAS4A& L1157mm & SVS13A & L1157-B1& L1448-R2& IRAS16293 & TAG \\ \hline
O- & CO    &    Y   &  Y        &  Y   &   Y   &   Y     &  Y     &   Y     &   Y     &   Y       & 28503 \\
   & HCO   &    Y   &  Y        &  Y   &   Y   &   Y     &  Y     &   Y     &   Y     &   Y       & 29004 \\
&  HCO$^+$ &    Y   &  Y        &  Y   &   Y   &   Y     &  Y     &   Y     &   Y     &   Y       & 29507 \\
& HOCO$^+$ &    Y   &  Y        &  Y   &   Y   &   Y     &  -     &   Y     &   Y     &   -       & 45010 \\
&H$_2$O    & (1)    &  -        & (2)  & ($\dag$) &   -  &($\dag$)& ($\dag$)& (3)     & ($\dag$)  & 19004 \\
&CCO       &    -   &  Y        &  -   &   -   &   -     &  -     &   -     &   -     &  -        & 40006 \\
& CCCO     &    Y   &  Y        &  Y   &   -   &   -     &  -     &   -     &   -     &   -       & 52501 \\
&{\bf CH$_3$OH}&Y   &  Y        &  Y   &   Y   &   Y     &  Y     &   Y     &   Y     &   Y       & 32003 \\
& H$_2$CO  &  $\dag$&  $\dag$   &$\dag$&   Y   &   (*)   &  Y     &   Y     &   (*)   &   Y       & 30501  \\
& H$_2$COH$^+$&  Y   &  (4)     &  -   &   Y   &   -     &  -     &   Y     &   -     &   -       & 31504 \\
& H$_2$CCO  &    Y   &  Y        &  Y   &   Y   &   Y     &  -     &   Y     &   Y     &   Y       & 42501 \\
& HCOOH     &    Y   &  Y        &  -   &   Y   &   Y     &  -     &   Y     &   Y     &   Y       & 46005 \\
& {\bf C$_2$H$_5$OH}   &- &  -      &  -   &   Y   &   -     &  -     &   Y     &   -     &   -       & 46004 \\
& {\bf HCCCHO}&    Y   &  Y      &  Y   &   Y   &   -     &  -     &   -     &   -     &   -      & 54007 \\
& {\bf CH$_3$CHO} &    Y   &  Y        &  Y   &   Y   &   Y     &  Y     &   Y     &   Y     &   Y       & 44003 \\
& {\bf HCOOCH$_3$} &   Y   &  Y        &  -   &   Y   &   -     &  Y     &   Y     &   -     &   Y       & 60003 \\
& {\bf CH$_3$OCH$_3$}& Y   &  Y        &  -   &   Y   &   -     &  Y     &   Y     &   -     &   Y       & 46514 \\
& {\bf HCOCH$_2$OH}&  -    &  -        & -   &   Y   &   -     &  -     &   Y     &   -     &   -       & 60501 \\  \hline
N- & CN        & $\dag$  &  Y        &  (*) &   Y   &   (*)   &  Y     &   Y     &   Y     &   Y       & 26001 \\
& HCN       &   Y    &  Y        &  Y   &   Y   &   Y     &  Y     &   Y     &   Y     &   Y       & 27501 \\
& HNC       &   Y    &  Y        &  Y   &   Y   &   Y     &  Y     &   Y     &   Y     &   Y       & 27002 \\
& HCNH$^+$  &   Y    &  (*)       &  (*) &   Y   &   (*)   &  -     &   -     &  (*)    &   -       & 28504 \\
& HNO       &   Y    &  Y        &  Y   &   -   &   -     &  -     &   Y     &   -     &   Y       & 31005 \\
& NH$_3$($\dag$)&  Y &  Y        &  Y   &   Y   &   Y     &  Y     &   Y     &   Y     &   Y       & 18004 \\
& N$_2$H$^+$&   Y    &  Y        &  Y   &   Y   &   Y     &  Y     &   Y     &   Y     &   Y       & 29506 \\
& HC$_3$N   &   Y    &  Y        &  Y   &   Y   &   Y     &  Y     &   Y     &   Y     &   Y       & 51501 \\
& HCCNC     &   Y    &  Y        &  Y   &   Y   &   -     &  Y     &   -     &   Y     &   Y       & 51004 \\
& HNCCC     &   Y    &  -        &  Y   &   -   &   -     &  -     &   -     &   -     &   -       & \\
& HNCO      &   Y    &  Y        &  Y   &   Y   &   Y     &  Y     &   Y     &   Y     &   Y       & 43002 \\
& HCNO      &   Y    &  Y        &  Y   &   Y   &   Y     &  -     &   Y     &   -     &   -      & 43509 \\
& HOCN      &   Y    &  Y        &  Y   &   Y   &   Y     &  -     &   -     &   -     &   -       & 43510 \\
& HSCN      &   -    &  Y        &  Y   &   Y   &   -     &  -     &   -     &   -     &   -       & 59505 \\
& {\bf HC$_3$NH$^+$}& Y &  Y     &  -   &   -   &   -     &  -     &   -     &   -     &   -       & 52503 \\
& {\bf HC$_5$N}& Y   & -         &  Y   &   -   &   Y     &  Y     &   Y     &   Y     &   Y      & 75503 \\
& CH$_2$CN  &   Y    &  Y        &  Y   &   -   &   -     &  Y     &   Y     &   -     &   -       & 40505 \\
& {\bf CH$_3$CN}& Y  &  Y        &  Y   &   Y   &   Y     &  Y     &   Y     &   Y     &   Y       & 41505 \\
& {\bf C$_2$H$_3$CN}& Y &  -     &  Y   &   -   &   -     &  -     &   Y     &   -     &   -       & 53515 \\
& {\bf NH$_2$CHO}& - &  -        &  -   &   Y   &   -     &  Y     &   Y     &   Y     &   Y       & 45512 \\
& CCCN      &   Y    &  Y        &  Y   &   -   &   -     &  -     &   Y     &   Y     &   -       & 50511 \\
 \hline
C- &CCH       &   Y    &  Y        &  Y   &   Y   &   Y     &  Y     &   Y     &   Y     &   Y       & 25501 \\
& c-C$_3$H  &   Y    &  Y        &  Y   &   Y   &   Y     &  -     &   -     &   Y     &   Y       & 37003 \\
& l-C$_3$H  &   Y    &  Y        &  Y   &   Y   &   Y     &  -     &   -     &   Y     &   Y       & 37501 \\
& c-C$_3$H$_2$& Y    &  Y        &  Y   &   Y   &   Y     &  Y     &   Y     &   Y     &   Y       & 38002 \\
& l-C$_3$H$_2$& Y    &  Y        &  Y   &   -   &   Y     &  -     &   -     &   -     &   -       & 38501 \\
& l-C$_4$H$_2$& Y    &  Y        &  Y   &   -   &   Y     &  -     &   -     &   Y     &   -       & 50503 \\
& C$_4$H    &   Y    &  Y        &  Y   &   Y   &   Y     &  Y     &   Y     &   Y     &   Y       & 49503 \\
& C$_5$H    &   -    &  -        &  Y   &   -   &   -     &  -     &   -     &   -     &   -       & 61003 \\
& C$_6$H    &   -    &  -        &  Y   &   -   &   -     &  -     &   -     &   -     &   -       & 73001 \\
& CH$_3$CCH &   Y    &  Y        &  Y   &   Y   &   Y     &  Y     &   Y     &   Y     &   Y       & 40001 \\
\hline
S- &CS        &   Y    &  Y        &  Y   &   Y   &   Y     &  Y     &   Y     &   Y     &   Y       & 44501 \\
& NS        &   Y    &  (*)      &  -   &   Y   &   Y     &  -     &   Y     &   Y     &   Y       & 46515 \\
& SO        &   Y    &  Y        &  Y   &   Y   &   Y     &  Y     &   Y     &   Y     &   Y       & 48501 \\
& SO$_2$    &   Y    &  Y        &  Y   &   Y   &   Y     &  Y     &   Y     &   Y     &   Y       & 64502 \\
& OCS       &   Y    &  Y        &  Y   &   Y   &   Y     &  Y     &   Y     &   Y     &   Y       & 60503  \\
& CCS       &   Y    &  Y        &  Y   &   Y   &   Y     &  Y     &   Y     &   Y     &   Y       & 56007 \\
& CCCS      &   Y    &  Y        &  Y   &   Y   &   Y     &  Y     &   -     &   Y     &   Y       & 68503 \\
& HCS$^+$   &   Y    &  Y        &  Y   &   Y   &   Y     &  Y     &   Y     &   Y     &   Y       & 45506 \\
& H$_2$CS   &   Y    &  Y        &  Y   &   Y   &   Y     &  Y     &   Y     &   Y     &   Y       & 46509 \\
& SO$^+$    &   -    &  Y        &  -   &   (*) &  (*)    &  (*)   &   Y     &   -     &   -       & 48010 \\
& {\bf CH$_3$SH} & - &  Y        &  -   &   Y   &   -     &  -     &   Y     &   -     &   Y       & 48510 \\
\hline
\end{tabular}
}
(1)~Caselli et al. (2012)
(2)~Karska et al. (2013)
(3)~Nisini et al. (2012)
(4)~Fuente et al. (2016)
\end{table*}

\begin{table*}
\centering{
\caption[]{3~mm band observations: Census of Silicon-bearing and Phosphorus-bearing molecular species detected. We have indicated the main isotopologues.  For each molecular species, we indicate the associated TAG, according to the database used for identification (CDMS or JPL).}
\begin{tabular}{llcccccccccc}
\hline
  &Source    &  L1544 & B1b & L1527& IRAS4A& L1157mm & SVS13A & L1157-B1& L1448-R2& IRAS16293 & TAG \\ \hline
Si- &SiO       &   -    &  Y        &  -   &   Y   &   Y     &  Y     &   Y     &   Y     &   Y       & 44505 \\
    &SiS       &   -    &  -        &  -   &   -   &   -     &  -     &   Y     &   -     &   Y       & 60506 \\ \hline
P-  &PN        &   -    &  Y        &  -   &   Y   &   -     &  -     &   Y     &   -     &   Y       & 45511 \\
    & PO       &   -    &  -        &  -   &   -   &   -     &  -     &   Y     &   -     &   -       & 47507  \\  \hline
\end{tabular}
}
\end{table*}

We have classified the identified molecular species in "elemental families": carbon, nitrogen, oxygen, sulfur, silicon, phosphorus bearing.
\begin{itemize}
\item The carbon family contains only hydrocarbons C$_x$H$_y$.
\item The oxygen family  contains only molecules of the type C$_x$H$_y$O$_z$.
\item The nitrogen family contains only molecules of the type C$_x$H$_y$O$_z$N$_t$.
\item The sulfur family contains any Sulfur-bearing molecules, of the type C$_x$H$_y$O$_z$N$_t$S$_u$
\item The silicon (phosphorus) family contains any Si-bearing (P-bearing) species.
\end{itemize}
These definitions are of course incomplete from a chemical point of view, when considering the molecular reaction network associated with each elemental family. A simple molecule like SO is obviously involved in both the sulfur and oxygen chemical networks. The above definitions are introduced for the sake of the analysis which follows hereafter, as they  permit identification of  evolutionary trends in the gas chemical composition.

Some  molecular species were identified from their rare isotopomers (D, $^{13}$C): they are indicated with a  $\dag$  in Table~2. In a few cases, species missed by ASAI in the 3mm band were known to be present  thanks to their signature in the 2~mm and 1.3~mm bands, like SO$^+$ or H$_2$CO. These few species are marked with an asterisk $(*)$ in Table~2. In the case of water, we have mentioned the detection of the main isotopologue by {\em Herschel}
when the rare isotopomers HDO and H$_2^{18}$O were not detected by ASAI. This is the case for L1544 (Caselli et al. 2012), L1527 (Karska et al. 2013) and L1448-R2 (Nisini et al. 2012). We also note that H$_2$COH$^+$ was marginally detected
towards B1b by Fuente et al. (2016), thanks to higher sensitivity observations. These cases are rare enough that they should not significantly affect the statistics nor our conclusions. Of course, the difference of main beam filling factors between extended sources (prestellar cores, shocks) and compact, strongly diluted hot corino sources, influence to some extent the conclusions of this analysis and this study will be extended to the analysis of the 2~mm and 1.3~mm, for which such a bias effect is less pronounced. On the other hand, the spatial distribution of the molecular emission also conveys some important information on the source chemistry.

The new generation of interferometers such as ALMA and NOEMA  reveals the presence at a much smaller angular scale of additional molecular species, which are missed by  sensitive single-dish surveys such as ASAI.  So far, only IRAS16293-2422 has been investigated in an unbiased manner in the course of the PILS survey (Jorgensen et al. 2016). Thanks to ALMA, spectral signatures of glycolaldehyde CH$_3$OHCHO (Jorgensen et al. 2012), acetic acid CH$_3$COOH,  ethylene glycol (CH$_2$OH)$_2$ (J{\o}rgensen et al. 2016) and  ethanol C$_2$H$_5$OH (Lykke et al. 2017) have been detected towards IRAS16293-2422. Some of the of the ASAI sources have been studied at high angular resolution (and high sensitivity with ALMA and NOEMA (see e.g. L\'opez-Sepulcre et al. 2017), but there is no coherent data set in terms of frequency range, sensitivity and angular resolution for the whole ASAI source sample.

We first made an inventory of the number of molecular species, counting separately the main isotopologues and the rare isotopologues detected. The detection of the latter crucially depends on the sensitivity of the survey and the gas column density. Hence, the number of detected isotopologues and molecular lines is a better indicator of the excitation conditions, whereas the number of main isotopologues rather traces the molecular complexity of the source. The results  are summarized in Tables~2 and 3, with the exception of TMC1, which was not observed at 3~mm. For the sake of comparison, we have included IRAS16293-2422, the only hot corino source investigated in detail until now (Caux et al. 2011; Jaber et al. 2014, 2017; Jorgensen et al. 2016). In order to avoid possible bias in the analysis, we take into account only the molecular transitions detected at an  angular scale  comparable to that of the ASAI survey ($10\arcsec$--$30\arcsec$) and in the same spectral range.

\begin{figure*}
\center{
\caption[]{Distribution of the composition of ASAI sources in chemical families, based on the number of molecular lines in the 3~mm band: Oxygen, Nitrogen, Carbon, Sulfur. The early prestellar  core TMC1  (observed only at 2mm) was left aside. We show the composition of the Class 0 source IRAS16293-2422 for the sake of comparison (Caux et al. 2011). Comparison of the C-rich and O-rich line number permits identification of the WCCC ($O/C\leq 1.5$) and hot corino ($O/C \gg 1.5$) families.   }
\includegraphics[width=2.5\columnwidth]{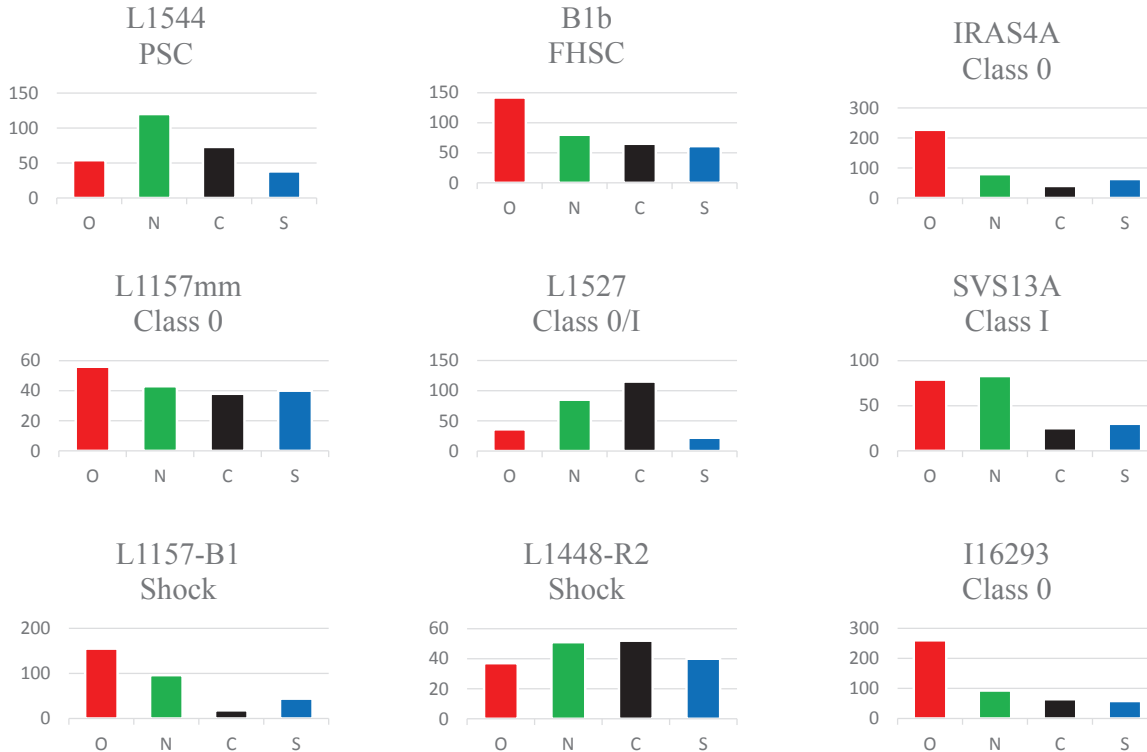}
}
\end{figure*}

The chemical richness can be first described by the number of detected main isotopologues in the ASAI sources, as shown in Table~5. It is typically 40, ranging from 35 (SVS13A) to 51 (B1b). This number varies little between sources, independently of their evolutionary stage and luminosity. Overall, there is no marked difference between the relative importance of the different chemical families.

The ratio $r$ of O-bearing/hydrocarbon {\em species} allows us to discriminate between the hot corino sources  ($r\simeq$ 1.9--3.8) and the WCCC sources ($r\simeq$ 1.0--1.4). The definitions adopted for the O- and C-families are very conservative but offers the advantage of lifting any ambiguity on the value of $r$ and $r^*$ (see below). We observe that the ratio $r$ tends to increase with time, from FHSC  ($r=1.9$) to Class I ($r=2.3$). Based on this ratio, we could classify L1157mm as a WCCC source ($r$= 1.4) and B1b and SVS13A as hot corino sources ($r$=1.9 and 2.3, respectively),  in agreement with the inventory of detected complex organic molecules (see Table 2). As can be seen in Table~5 and Fig.~4, the difference between O-rich and hydrocarbon-rich sources appears markedly when considering the ratio $r^{*}$ of the number of molecular {\em lines} from O- to C-bearing species.  The range of values is $r^{*}$= 0.3--1.5 for WCCC sources and $r^{*}$= 2.2--8.6 for hot corino sources.

A trend seems to emerge between the chemical properties of the  envelopes and the location of the protostars inside the molecular clouds. The large SCUBA survey of the Perseus molecular cloud by Hatchell et al. (2005) shows that the three hot corino sources of our sample (B1b, IRAS4A, SVS13A)  are located inside dense filamentary clouds (see also Savadoy et al. 2014). This is unlike
L1527, which is located outside the dense filaments of the Taurus molecular cloud (Goldsmith et al. 2008)
and L1157-mm, which lies in a small isolated cloud (see e.g. Chiang et al. 2010).  This suggests that
environmental conditions could play an important role in the origin of the chemical protostellar diversity.
Systematic studies on  a much larger sample should be carried out in order to confirm this trend.

Two ASAI  sources display a low content in Carbon-bearing species: SVS13A and L1157-B1. It is difficult to conclude if this is a consequence of specific chemical activity associated with outflows, or simply a time evolution effect, or both.

Since many sources are associated with outflow shock regions inside the telescope beam, it may seem not so surprising that many lines and species are detected in common towards protostars and pure outflow shock regions. However,
the two protostellar shocks studied here display unexpected results: L1157-B1 (in a WCCC protostellar region) has a signature typical of hot corino regions (r=3.8), a fact confirmed by the large number of detected COMs in the shock (Lefloch et al. 2017), whereas L1448-R2, associated with a hot corino region, has a signature typical of WCCC (r=1.3). Detailed analysis of the chemical conditions in L1448-R2 is currently going on to understand why so many C-chains are still present in the shock.

The number of detected S-bearing species is rather constant. CH$_3$SH is the only S-bearing COM in the ASAI sample.  D-isotopologues are detected at all stages of evolution. The number of deuterated species is found quite high in the prestellar and Class 0 phase, before dropping in the advanced Class I phase (Bianchi et al. 2017). Many deuterated species are  in both shocks, which reveal the presence of fossil deuteration in the outflowing gas (see Codella et al. 2012).

Overall, we detect from $\approx 200$  to 400 molecular lines per source in the 3~mm band. The number of detected molecular lines increases with source luminosity in the early protostellar stage, and decreases with evolutionary stage, from Class 0 to Class I. We do not observe any marked difference in terms of spectral line density between prestellar and protostellar objects, with $\sigma\simeq 6-12$ lines/GHz (Table 5). Unexpectedly, the spectral line density measured in protostellar shocks and prestellar cores is  comparable to that measured towards protostars: the lowest spectral line density are measured towards the lower luminosity protostars and/or more evolved protostars.

Interestingly enough, the number of main isotopologues detected towards the prototypical hot corino source IRAS16293-2422 (Caux et al. 2011) and the high-mass star forming region Orion KL (Tercero et al. 2010) are rather similar to those of the ASAI sources: 41 and 43, respectively.
A difference appears when considering the number of lines in the ASAI sources: the number of detected lines in the 3~mm band is typically 200--400, hence fully comparable to that obtained in Class 0 solar-type protostar IRAS16293-2422, but it is typically a factor of 10 times less than what is found towards Orion KL (Tercero et al. 2010). A similar behaviour was observed in the  submillimeter to far-infrared domain by Herschel (Ceccarelli et al. 2010), supporting the idea that this result is rather general and frequency independent.
Comparing the chemical composition of hot corino sources, only little variations are observed between B1b, IRAS4A, IRAS16293-2422, and SVS13A (Fig.~4). The main difference seems to arise from the content in N-bearing species, relatively more important in Class I SVS13A.

Our main conclusion is that chemical richness is already widely present in the initial phases of solar-type star formation, and it is not less than towards high-mass star forming regions. Prestellar cores present a high degree of molecular diversity, comparable to that of protostars. The number of detected molecular lines does not appear to depend much on the evolutionary stage of the source. The large difference in spectral line density appears as a consequence of the excitation conditions  and not major differences in the chemical richness. Of course, there are strong differences in the molecular abundances even if the species are detected in all sources. In this context, the long studied solar-type protostar IRAS16293-2422 does not appear to be a unique object, but rather a typical object of the hot corino Class 0, when compared with the other sources of the ASAI sample.

\subsection{Unidentified lines}

\begin{table*}
\centering{
\caption[]{Frequency and Observational parameters of U lines in the ASAI source sample.  The observational parameters (Flux, linewidth FWHM, peak flux $T_{A}^{*}$) were obtained from a simple gauss fit to the line profile. The uncertainties are given in brackets. Velocity-integrated fluxes and line intensities are expressed in units of antenna temperature corrected for atmospheric attenuation. Previously detected lines are commented. }
\begin{tabular}{lrrrrl}
\hline
Source & Rest Frequency  & Flux        &  FWHM & $T_{A}^{*}$ & Comment \\
       &   MHz      & (mK$\kms$)  & ($\kms$)  & (mK)     & \\ \hline
L1544  & 83289.27   & 7(1)        & 0.32(.05) & 20.1(2.8)&  \\  
       & 86001.00   & 6(1)        & 0.49(.09) & 11.7(2.3)&  \\   
       & 86978.65   & 8(1)        & 0.63(.11) & 13.6(2.6)& Previously seen by Jones et al. (2007) \\
\hline
B1b    & 81571.8    & 59(8)     & 1.28(.17) & 43.2(5.6)\\
       & 82395.0    & 96(9)     & 1.67(.17) & 53.9(5.3)\\
       & 91344.0    & 32(5)     & 3.18(.60) & 9.6(1.7)\\
       & 100198.5   & 56(3)     & 1.27(.09) & 42.1(2.2)& \\ \hline
L1527  & 91425.8    & 21(4)       & 1.1(.2)   & 19.5(3.5)& \\
       & 98832.9    & 26(5)       & 1.7(.4)   & 13.5(3.5)& \\ \hline
IRAS4A & 85294.8    & 20(4)       & 1.2(.3)   & 16.0(2.9)& \\
       & 90669.8    & 18(3)       & 1.6(.3)   & 11.0(1.8)& Previously seen by Ziurys et al. (1988)\\
       & 91359.0    & 38(4)       & 1.6(.1)   & 22.2(2.6)& \\
       &100198.5    & 25(2)       & 1.2(.1)   & 19.8(2.4)& \\
       &112563.3    & 40(8)       & 1.8(.4)   & 21.1(5.7)& \\ \hline  
L1157-mm   & 94500.0  & 18(4)       & 1.5(.4)   & 11.0(3.0)& Previously seen by Snyder et al. (2002) \\  
          &100198.5  & 17(3)       & 1.4(.3)   & 11.3(2.4)& \\
         &110575.9  & 66(14)      & 1.6(.4)   &40.0(9.8) & Previously seen by Combes et al. (1996) \\ \hline
SVS13A   & 93187.16 & 31(4)       & 1.4(0.2)  & 20.7(1.8) &  \\
         & 95849.59 & 17(3)       & 1.6(0.3)  & 9.9(2.1) &  \\ \hline
L1157-B1  &100198.5  & 46(6)       & 5.3(.8)   & 8.2(1.6) & \\
\hline
L1448-R2 & 89974.9  & 59(9)       & 4.8(.9)   & 11(3) & \\
         & 92374.0  & 27(6)       & 1.3(.3)   & 20(4) & \\ 
\hline
\end{tabular}
}
\end{table*}

For each source, a few lines detected at the $5\sigma$ level remain unidentified (U) in the 3~mm band. Other U lines are present in the spectral surveys, though at a lower level of detection
($3\sigma$ and $4\sigma$), which cast some doubts on their statistical significance. In order to provide a list of useful and reliable U lines for future spectroscopic work, we have chosen the $5\sigma$ level as detection criterion.

The number of U lines per source is low (Table 4) and represent 1\%--2\% of the number of detected lines (per source). In Table~5, we report the frequency and observational parameters of these U lines.  The reported frequency was chosen so that the emission peak velocity coincides with the $v_{lsr}$ of the source. The uncertainty on the frequency is typically 0.3 MHz ($\approx 1\kms$) for regions associated with velocity gradients, such as L1157-B1 or L1448-R2. Some U lines were known from previous works. We have indicated the original reference whenever the case in Table~4. There is little match between the list of U lines identified in the different sources, but this is mainly a consequence of our selection criterion, as we report here only the lines detected above the $5\sigma$ level.

In the case of L1157-B1, we detected only 1 line  without any proper identification from the JPL and CDMS public line catalogs. We do not confirm any of the U lines reported by Yamaguchi et al. (2012) in their NRO 45m survey of the shock.  The higher sensitivity of the IRAM 30m data, typically a factor of 5, gives us confidence in our conclusion.

The low number of U lines present in the 3~mm spectral band indicates that they are probably due to "small" molecules, (molecules with a low number of atoms) as opposed to COMs. The content of the 2~mm and 1.3~mm bands remains to be analyzed. Making use of the additional information contained in the  higher frequency bands, we hope that collaborations with experts in molecular spectroscopy will allow us to identify the carriers of these U lines.

This is illustrated by our recent identification of the U line at 100198.5 MHz, which  is detected in several sources: the shock region L1157-B1, the envelope of the driving protostar L1157-mm, the Class 0 protostar IRAS4A, and the FHSC B1b. This line is detected in other sources, like in L1544, with a lower SNR. Using complementary frequency laboratory  measurements, we could assign this line to the rotational transition J=2-1 of NS$^+$, and we could identify the additional transitions J=3-2 and J=5-4 in the ASAI spectra (Cernicharo et al. 2018).

\begin{table*}
\centering{
\caption[]{3~mm band observations: Number of molecular species (main isotopologue), molecular transitions and spectral line density $\sigma$ (in lines/GHz) for each source of the ASAI sample. We have included IRAS16293-2422 for comparison. O-COMs: CH$_3$OH, HCCCHO, HCOOCH$_3$, CH$_3$OCH$_3$, HCOCH$_2$OH; N-COMs: HC$_5$N, CH$_3$CN, C$_2$H$_3$CN, NH$_2$CHO}
\begin{tabular}{lrrrrrrrrr}
\hline
Source            &  L1544 & B1b & L1527 & IRAS4A & L1157mm & SVS13A & L1157-B1 & L1448-R2 & IRAS16293 \\
Type              &   PSC  &FHSC & {Class 0/I} & Class 0& Class 0 & Class I& Shock    & Shock    & Class 0 \\ \hline
O-                &    14  &  15&   10    &   16 &     10  &    9   &     15   &      9   &    11   \\
N-                &    19  &  16&   19    &   14 &     12  &   12   &     15   &     13   &    12    \\
C-                &     8  &   8&   10    &   6  &      7  &    4   &      4   &      7   &     6    \\
S-                &     9  &  10&    8    &   10 &     11  &    9   &      9   &      9   &     9    \\
Si-               &     0  &   1&    0    &   1  &      1  &    1   &      2   &      2   &     2    \\
P-                &     0  &   1&    0    &   1  &      0  &    0   &      2   &      0   &     1    \\
\hline
Total Species     &    50  &  51&   47    &   48 &    41   &    35  &     47   &     40   &    41  \\
\hline
D-                &   11   &  18&   11    &   11 &     7   &    3   &      3   &      8   &    13    \\
\hline
$r$ (O-/C-)       &   1.8  & 1.9&   1.0   &  2.7 &    1.4  &   2.3  &    3.8   &    1.3   &  1.8 \\
\hline
O-Lines           &   54   & 142&   36    &  227 &     56  &     79 &    155   &      37  &  260   \\
N-Lines           &  124   &  80&   85    &   79 &     43  &     83 &     96   &      51  &   93    \\
C-Lines           &   73   &  65&   115   &   40 &     38  &     25 &     18   &      52  &   64    \\
S-Lines           &   38   &  61&    22   &   63 &     40  &     30 &     44   &      40  &   58    \\
Si-Lines          &    0   &   1&     0   &    3 &      1  &      1 &      6   &       4  &    3    \\
P-Lines           &    0   &   1&     0   &    1 &      0  &      0 &      5   &       0  &    1    \\
\hline
Total Lines       & 289    & 350&   258   &  413 &    178  &    218 &    324   &     184  &  479     \\
\hline
D-Lines           &  47    &  62&    46   &   75 &     31  &     15 &      5   &      30  &   79   \\
\hline
$r^*$ (O-/C-lines)&   0.7  & 2.2&   0.3   &  5.7 &    1.5  &    3.2 &    8.6   &     0.7  &  4.1   \\
\hline
O-COMs Lines      &  35    & 109&    14   &  197 &     36  &     63 &    139   &      27  &  244  \\
N-COMs Lines      &  32    &  7 &    16   &   23 &      7  &     21 &     46   &      10  &   38  \\
\hline
U Lines           &  3     &  4 &   2     &    5 &      3  &      2 &      1   &       2  &   - \\
U Lines/Total Lines (\%)      &  1.0   & 1.2&   0.8   &  1.2 &     1.7 &     0.9&    0.3   &     1.0  &   - \\
\hline
$\sigma$ (Lines/GHz)& 8.9  &11.0&   8.0   & 11.8 &    5.3  &    6.1  &   9.1   &     5.8  & 13.3    \\
\hline
\end{tabular}
}
\end{table*}

\section{Chemical evolution along star formation}
From our results it emerges a relatively simple picture of the chemical evolution during the star formation  process, which will be depicted here.

\subsection{Prestellar cores}
Observations of the molecular composition of pre-stellar cores, the simplest sites where solar-type stars form, have revealed a very systematic pattern of chemical differentiation (Bergin \& Tafalla, 2007; Ceccarelli et al. 2007; Caselli \& Ceccarelli 2012). During the cold and dense pre-collapse phase, molecules
freeze-out onto the grain surfaces, forming ices. Subsequent hydrogenation of atoms and CO on the grain
surface leads to the formation of more complex organic molecules, like  formaldehyde (H$_2$CO)
and methanol (CH$_3$OH), in addition to other hydrogenated species (Watanabe et al. 2003, 2007; Fuchs et al. 2009; Pirim et al. 2010; Taquet et al. 2012; Rimola et al. 2014).

The first results from ASAI reveal that even the dense cores, which are usually assumed to be particularly simple, are much more complex and chemically rich than previously thought.
The spectrum of the prestellar cores is particularly rich in the 3~mm band, and it is dominated by carbon-containing molecules. From the L1544 spectral scan we have been able to obtain a full census of the oxygen-bearing COMs in this source. We have detected tricarbon monoxide (C$_3$O), methanol (CH$_3$OH), acetaldehyde (CH$_3$CHO), formic acid (HCOOH), ketene (H$_2$CCO), and propyne (CH$_3$CCH) with abundances varying from 5 $\times$ 10$^{-11}$ to 6 $\times$ 10$^{-9}$ (Vastel et al. 2014); it was found that a non-thermal desorption mechanism is possibly responsible for the observed emission of methanol and COMs from the same external layer. Subsequent targeted observations have confirmed the presence of a variety of COMs, including methyl formate (HCOOCH$_3$) and dimethyl ether CH$_3$OCH$_3$ (Jimenez-Serra et al. 2016).  Methanol maps of L1544 by Bizzocchi et al. (2014) confirmed that the emission mainly arises from the outer parts of the core, where CO just started to freeze-out onto dust grains. Similar results on the presence of COMs in dark clouds and prestellar cores were obtained by other groups approximately at the same time (\"{O}berg et al. 2010,  Cernicharo et al. 2012, Bacmann et al. 2012). Based on the WHISPER survey of the Horsehead nebula and a nearby dark core,  Guzman et al. (2014) reached the same conclusion about the origin of CH$_3$OH emission: it is present mainly in the envelope. The authors conclude that a pure gas phase model can account for the observed abundance of H$_2$CO.

The ASAI detection of the cyanomethyl radical (CH$_2$CN) for the first time in a prototypical prestellar core (L1544) was reported by Vastel et al. (2015b). The authors were able to identify the hyperfine transitions of the ortho and para forms and computed all transition frequencies and line intensities for all transitions including satellite hyperfine components at the frequencies observed by ASAI. That paper also reported the first detection of the fine and hyperfine structure of the ortho and para forms of the cyanomethyl radical at ~101 GHz, resolved in this cold dense core.

Several molecular ions have been identified.  Lines from the protonated carbon dioxide ion, HOCO$^+$ were analysed under non-LTE assumptions showing  that the HOCO$^+$ emission originates in the same external layer where non-thermal desorption of the other species mentioned above has been observed. Its abundance relative to \htwo\ is found to be  (5 $\pm$ 2) $\times$ 10$^{-11}$, and pure gas phase models of the chemistry involved in the formation and destruction of HOCO$^+$ provide a gaseous CO$_2$ abundance of 2 $\times$ 10$^{-7}$ (with respect to H$_2$) with an upper limit of 2 $\times$ 10$^{-6}$ (Vastel et al. 2016).
Nitrogen ions such as HCNH$^+$ and HC$_3$NH$^+$ have also been detected for the first time in a prestellar core (Qu\'enard et al. 2017). The high spectral resolution of the observations allows to resolve the hyperfine structure of HCNH$^+$. A radiative transfer modelling of these ions leads to abundances of $3\times$ 10$^{-10}$ for HCNH$^+$ and  $(1.5-3)\times 10^{-12}$ for HC$_3$NH$^+$, with respect to H$_2$. The study of some nitrogen species linked to their production (HCN, HNC, HC$_3$N)  coupled with a gas-grain chemical modelling shows that the emission of these ions originates in the external layer where non-thermal desorption of other species was previously observed.

\subsection{Protostars}

The spectrum of Barnard~1, a protostellar object intermediate between prestellar cores and Class 0 sources, is characterized by many lines of deuterated species, complex molecules and Sulfur-bearing molecules. Fuente et al. (2016) provided a very complete inventory of neutral and ionic C-, N- and S- bearing species with, including the detections of HOCO$^+$ and HCNH$^+$, a tentative detection of HC$_3$NH$^+$, and up to our knowledge, the first secure detections of the deuterated ions DCS$^+$ and DOCO$^+$. The authors also determined the value of the cosmic ray ionization rate and the depletion factors. The observational data were well fitted with $\zeta_{H_2}$ between 3 $\times$ 10$^{-17}$ s$^{-1}$ and 10$^{-16}$ s$^{-1}$. Elemental depletions were estimated to be $\sim$10 for C and O, $\sim$1 for N and $\sim$25 for S. B1b presents similar depletions of C and O than those measured in pre-stellar cores. The depletion of Sulfur was found to be higher than that of C and O but not as extreme as in cold cores. In fact, it is similar to the values found in some bipolar outflows, hot cores and photon-dominated regions, which could be the consequence of the initial conditions (important outflows and enhanced UV fields in the surroundings) and a rapid collapse ($\sim$0.1 Myr) that permits to maintain most S- and N-bearing species in gas phase to great optical depths.

The line emission of c-C$_3$H$_2$ and its $^{13}$C isotopic species were found particularly interesting in the WCCC source L1527, so we conducted a study which confirms the dilution of the $^{13}$C species in carbon-chain molecules and their related molecules (Yoshida et al. 2015). The rare isotopologues are thought to originate from
$^{13}$C$^+$, which is  diluted  in the gas-phase  due to the isotope exchange reaction:  $^{13}$C$^{+}$ + $^{12}$CO $\rightarrow$ $^{12}$C$^{+}$ + $^{13}$CO. This exchange reaction is exothermic ($\Delta E$= $35\K$) so that the $^{12}$C$^+$/$^{13}$C$^+$ ratio tends to be higher in cold clouds and  the  $^{13}$C species of various molecules produced from $^{13}$C$^+$ become less abundant.  Moreover, the abundances of the two $^{13}$C isotopic species are different from each other. The ratio of c-$^{13}$CCCH$_2$ species relative to c-CC$^{13}$CH$_2$ is determined to be 0.20 $\pm$ 0.05.
If $^{13}$C were randomly substituted for the three carbon atoms, this ratio would be 0.5.
Hence, the observed ratio indicates that c-$^{13}$CCCH$_2$ is favoured in dense cores.

The high-sensitivity spectrum obtained towards the Class 0 source L1157-mm looks very similar to that of L1527, and does not display the COM-rich spectrum of objects like IRAS4A (see Figs.1--3). This leads to classify L1157-mm as another member of the WCCC class.  The long carbon-chains C$_5$H and C$_6$H were not detected in the protostellar envelope of L1157-mm. It would be useful to carry out  complementary observations in the low frequency range, in which their rotational transitions are more easily detected (see e.g. Sakai et al. 2008), in order  to confirm their presence or not. The L1157-mm spectrum also reveals blue and red detached components about 45 km s$^{-1}$ away from the ambient cloud in the profiles of the SiO(5-4) and SiO(6-5) lines (Tafalla et al. 2015). These extremely high-velocity (EHV) components are similar to those found in the L1448 and IRAS 04166+2706 outflows and probably arise from a molecular jet driven by L1157-mm that excites L1157-B1 and the other chemically active spots of the L1157 outflow. This jet was recently mapped with NOEMA by Podio et al. (2016).

A very interesting result from ASAI is that the phase of hot corino can persist until the phase of Class I, as illustrated by the observations of SVS13A (Codella et al. 2016). In the spectrum of SVS13A, we clearly detected 6 broad (FWHM $\sim$ 4-5 km/s) emission lines of HDO with upper level energies up to E$_u$ = 837 K (Codella et al. 2016). A non-LTE LVG analysis implies the presence of very hot (150-260 K) and dense (> 3 10$^{7}$ cm$^{-3}$) gas inside a small radius ($\sim$ 25 AU) around the star, which is a clear indication, for the first time, of the occurrence of a hot corino around a Class I protostar. Although the effects of shocks and/or winds at such small scales can not be excluded, this could imply that the observed HDO emission is tracing the water abundance jump expected at temperatures $\sim$ 220-250 K, when the activation barrier of the gas phase reactions leading to the formation of water can be overcome. We derive X(HDO) $\sim$ 3 10$^{-6}$, and a H$_2$O deuteration > 1.5 10$^{-2}$.
The presence of a hot corino was confirmed by Bianchi et al. (2017), who showed evidence for a methanol enriched small region (radius $\simeq 35$ AU) of  hot (T$\sim 80\K$) and dense gas ($n(\htwo) \geq 10^8\cmmt$) around the protostar. More importantly, the analysis of the emission of the deuterated isotopologues of H$_2$CO and CH$_3$OH, showed that the deuteration richness drops in the Class I phase.
Recent observations with the Plateau de Bure Interferometer at $0.3\arcsec$ showed that the hot corino emission actually arises from component VLA4B of the binary (Lef\`evre et al. 2017).

\subsection{Protoplanetary phase}
In the subsequent phases (Class I/II), the envelope dissipates as the matter accretes onto the central
object and is dispersed by the outflow/jet, as the surrounding protoplanetary disk becomes detectable.
Chemistry is expected to be somewhat similar to that of Class 0 sources, with a corino region where ices
sublimate, and to the prestellar cores, in the cold and dense regions close to the equatorial plane. 

The spectral line density of AB Aur is very low, as expected in a protoplanetary disk. Nevertheless, several lines of CO and its isotopologues, HCO$^+$, H$_2$CO, HCN, CN, and CS,
were detected (Pacheco-V\'asquez et al. 2015). In addition, the detection of the SO $5_4$--$3_3$ and $5_6$--$4_5$ lines confirms the previously tentative detection of Fuente et al. (2010), which
makes AB Aur the only protoplanetary disk detected in SO. This detection indicates that this disk is warmer and younger than those associated with T Tauri stars. These ASAI results prompted follow-up studies at very high resolution which revealed the detailed structure of the AB Aur disk (Pacheco-V\'asquez et al. 2016).

\subsection{Outflow shocks}
Simultaneously with matter accretion onto the protostar, fast jets, possibly surrounded by a wider angle wind, are powered by the nascent star and seen to interact with the parental medium through molecular bowshocks, producing a slower moving molecular "outflow cavity" (Bachiller 1996). Outflow shocks compress and heat the interstellar material and grain ice mantles are sputtered, resulting in an especially rich chemistry (Bachiller et al. 2001; Codella et al. 2010). The extremely-high-velocity (EHV) gas forming "molecular bullets" is well differentiated from the gas traced by the standard outflow cavity wings and could represent material directly ejected from the prostostar or its immediate vicinity (Tafalla \& Bachiller 2011). Outflows contribute to
dissipate the circumstellar envelope, permitting the radiation of the central object to escape at increasing
distances, until the central star and its surrounding protoplanetary disk become optically visible as a (Class II) T Tauri star.

The L1157-B1 hot spot exhibits a particularly rich spectrum with hundreds of lines of many different molecular species (see Table 2), which makes it an ideal laboratory not only for shock but also for astrochemical studies.
Most of the lines tentatively detected by the NRO 45m telescope  could be confirmed by ASAI (see Table 3 in Yamaguchi et al. 2012), in particular the tentative detections of Si$^{18}$O and SiS. There are two exceptions however: first, Lefloch et al. (2017) assigned the transition CH$_3$OCHO 7(4,4,1)-6(4,3,1) to the line at 86224.16 MHz, instead of (CH$_3$)$_2$O at 86223.78 MHz; second, Mendoza et al.(2014) assigned the transitions HCNO 4-3 to the line at 91751.32 MHz, instead of HSCN 8(0,8)-7(0,7), in agreement with the detection of HCNO 5-4 at 114688.38 MHz (Mendoza et al. 2014). Several species are now detected, including complex N-bearing species  (C$_2$H$_3$CN, CH$_2$CN), long carbon-chain molecules (C$_4$H, HC$_5$N), and deuterated molecules like HDCS, NH$_2$D or HDO (see also Codella et al. 2012).

Podio et al. (2014) presented a complete census of molecular ions with an abundance greater than $\sim$10$^{-10}$, reporting the detection of  HCO$^+$, H$^{13}$CO$^+$, N$_2$H$^+$, HCS$^+$, and for the first time in a shock, from HOCO$^+$ and SO$^+$. It was concluded that HCO$^+$ and N$_2$H$^+$ are a fossil record of pre-shock gas in the outflow cavity, whilst HOCO$^+$, SO$^+$, and HCS$^+$ are effective shock tracers that can be used to infer the amount of CO$_2$ and sulphur-bearing species released from dust mantles in the shock.
The work on sulphur bearing ions predicted OCS to be a major carrier of sulphur on dust mantles and modelling of
H$_2$S observations suggest H$_2$S should not be a major carrier of sulphur on the grains (Holdship et al. 2016).

A multiline analysis of CS and its isotopic variations allowed to study the density structure of the molecular outflow (Gómez-Ruiz et al. 2015). The line profiles can be well fitted by a combination of two exponential laws that are remarkably similar to what previously found using CO by Lefloch et al. (2012). The CS observations show that this molecule is highlighting the dense, n$_{H_2}$ = 1-5 $\times$ 10$^{5}$ cm$^{-3}$, cavity walls produced by the episodic outflow in L1157. In addition, the highest excitation (E$_u$ $\geq$ 130 K) CS lines provide us with the signature of denser (1-5 $\times$ 10$^{6}$ cm$^{-3}$) gas, associated with a molecular reformation zone of a dissociative J-type shock, which is expected to arise where the precessing jet impacting the molecular cavities. The CS fractional abundance increases up to 10$^{-7}$ in all the kinematical components. This value is consistent with what previously found for prototypical protostars and it is in agreement with the prediction of the abundances obtained via chemical modelling.

Podio et al (2017) reported the detection of SiS for the first time in a low-mass forming region, in particular the protostellar shock L1157-B1. Complementary observations of the shock obtained with the Plateau de Bure Interferometer show for the first time that SiO and SiS have a different spatial distribution. Whereas SiO is a well established probe of silicates released from dust grain in shocks (see e.g. Gusdorf et al. 2008), SiS appears as a product of gas phase reactions between species released from dust grains in the shock. There is currently no satisfying formation scenario and more work is needed to understand the chemistry of Si-bearing molecules.

Thanks to the high sensitivity of the ASAI observations and full coverage of the 1.3~mm, 2~mm and 3~mm wavelength bands, a systematic search by Lefloch et al. (2017) led to the unambiguous detection of several COMs, such as ketene (H$_2$CCO), dimethyl ether (CH$_3$OCH$_3$) and glycolaldehyde (HCOCH$_2$OH) for the first time, and other tentative detections (Arce et al. 2008) were firmly confirmed, such as formic acid (HCOOH) and ethanol (C$_2$H$_5$OH).  The large number of detected lines from each species ($\sim$10--125) permitted accurate determination of their excitation conditions and molecular abundances. Combining these  results with those previously obtained towards other protostellar objects,  a good correlation is observed between ethanol, methanol and glycolaldehyde. The COM richness of the shock region L1157-B1 is fully comparable to that observed in hot corinos, both in terms of molecular abundances and chemical diversity.  The potential of shocks to study prebiotic chemistry is illustrated by the detection of formamide (NH$_2$CHO) in L1157-B1 by Mendoza et al. (2014).
This species is detected in a wide range of Galactic and Extragalactic environments. It turns out that the abundance in  the L1157-B1 shock region, $\approx 5\times 10^{-9}$, is among the highest values reported.

\subsection{Pre-biotic molecules along protostellar evolution}
L\'opez-Sepulcre et al. (2015) have conducted some first comparative studies for molecular species of particular interest, such as isocyanic acid (HNCO) and formamide. HNCO was detected in all the ASAI targets, except AB Aur, and NH$_2$CHO in five of them. Their abundances  was derived and analysed them together with those reported in the literature for high-mass sources. For those sources with formamide detection, a tight and almost linear correlation was found between HNCO and NH$_2$CHO abundances, with their ratio being roughly constant, suggesting that the two species are chemically related. The sources without formamide detection, which are also the coldest and devoid of hot corinos, fall well off the correlation, displaying a much larger amount of HNCO relative to NH$_2$CHO.  These results suggest that HNCO can be formed in the gas-phase during the cold stages of star formation. On the contrary NH$_2$CHO forms preferentially in warm environments. Recent theoretical and observational studies (Barone et al. 2015, Skouteris et al. 2017) suggest that formamide could form efficiently in the gas phase from the reaction between NH$_2$ and H$_2$CO. The tight correlation observed between formamide and isocyanic acid remains to be understood.

Formamide is one example of molecules of pre-biotic interest, which ASAI has allowed us to search for in solar-type star forming regions. Another example is provided by the search for Phosphorus-bearing species. Despite a low elemental abundance ($3\times 10^{-7}$, Asplund et al. 2009), phosphorus plays an important role in biochemical and metabolic processes in living beings. The signature of PN was detected in three sources of ASAI, two of which are protostars (see Table 3). For the first time, PO was detected in a solar-type protostellar shock region (Lefloch et al. 2016). Phosphorus-bearing species have been detected mainly in high-excitation shock regions until now (see also Ziurys 1987), similarly to Silicon-bearing species. Their non-detection towards the prestellar core L1544 is therefore not surprising. The lack of detection towards
L1527 or SVS13A is consistent with shocks of low excitation in the envelope.

\section{Conclusion}
We have undertaken the Large Program ASAI  to characterize and to understand the chemical evolution along solar-type protostellar evolution. To do so, we  have used the IRAM 30m telescope to carry out unbiased millimeter line surveys between 80 and 272 GHz of a sample of ten template sources, which cover the full formation process of solar-type stars, from prestellar cores to protoplanetary disks. At the time of writing this paper, the ASAI data analysis is under way and more results on the different sources and transversal studies will be published in forthcoming papers.
Below, we summarize our first conclusions based on the analysis of the 3~mm data: \\
\begin{itemize}
\item Molecular complexity is already present at early prestellar and protostellar stages. The chemical richness is already comparable to that of massive hot core regions, like Orion KL;
\item The number of detected molecular species varies little between sources, independently of their evolutionary stage and their luminosity. The number of detected molecular lines increases with source luminosity in the early protostellar stage, and decreases with evolutionary stage, from Class 0 to Class I. We do not observe any marked difference in terms of spectral line density between prestellar and protostellar phase. Unexpectedly, the spectral line density measured in protostellar shocks is fully comparable to that measured towards protostars;
\item The ratio of O- to C-bearing species/lines allows to identify very well the hot corino from the WCCC sources. This has permitted identification of a new WCCC (L1157-mm) and hot corino (SVS13A, B1b) sources.
\item  The hot corino sources of our sample are located inside dense filamentary clouds, whereas the WCCC sources are located outside dense filamentary structures.
\item The census of deuterated isotopologues shows similar number of detected species in the prestellar and early (Class~0) protostellar phase, independently of the chemical protostellar type (hot corino vs WCCC).
\item Our analysis provides a first glimpse on the prebiotic chemistry at work in solar-type star forming regions. Formamide, P-bearing species and glycolaldehyde are detected species of high interest, which deserve further studies.
\item COMs are commonly found at all stages of protostellar evolution, including energetic environments such as shocks.
\item A few U lines, whose number amount to 1\%--2\% of the total number of detected lines, remain to be identified in the 3~mm band for each source. Such low numbers  of U lines imply that we have obtained a comprehensive census of the COMs present in the sources of the sample.
\end{itemize}

From this preliminary analysis, a relatively detailed picture of the chemical evolution from prestellar cores to evolved Class II protostars can be drawn, while opening new questions, which will be addressed in further studies. It is important to confirm whether the difference of environmental conditions between hot corino and WCCC sources (inside/outside dense filamentary cloud regions) is a general trend. If so, it would provide important constraints on the origin of the chemical diversity observed in protostellar envelopes. As a consequence of the limited angular resolution of the survey ($\approx 20\arcsec$ in the 3mm band), several physical components may be contributing the detected emission: shocks, disk, hot corino, cold envelope. Source multiplicity (like in IRAS4A, B1b or SVS13A) makes it more difficult to interpret the results on the overall chemical evolution of the sources, especially in the case where the sources are  at different evolutionary stages.
The question as to whether chemical complexity increases, how much it differs in the close environment of protostars, the chemical processing associated with each physical component has to be  addressed  with higher angular resolution observations. It is one of the goals of the IRAM/NOEMA Large Program "Seeds Of Life In Space" (SOLIS\footnote{http://ipag.osug.fr/solis}; Ceccarelli et al. 2017).

Our results also illustrate the critical need of combining laboratory measurements and astronomical observations, as well as the need for a re-evaluation of the chemical processes at work in the gas phase and the characterization of molecular species neglected until now. We also stress the need for the computation of accurate collisional excitation rate coefficients.

In many cases, collisional rates with \htwo\ are simply not available. As an alternative, collisional rates with He (when available) or more simple hypothesis like LTE are used in order to determine molecular column densities, which therefore suffer some ambiguities in their determination. We hope that the wealth of molecular species commonly observed in the ASAI database will encourage experts in collisional rate computation to pursue and increase their efforts in the domain.

\section*{acknowledgements}
We thank an anonymous referee for useful comments and suggestions.
We thank Dr. A. Faure for discussions on Nitrogen isotopologues.
Based on observations carried out as part of  the Large Program ASAI (project number 012-12) with the IRAM 30m telescope. IRAM is supported by INSU/CNRS (France), MPG (Germany) and IGN (Spain). This work was supported by the Programme National "Physique et Chimie du Milieu Interstellaire" (PCMI) of CNRS/INSU with INP/INC, co-funded by CNES and CEA. J.C. and N.M. thanks Spanish MINECO for funding under grants AYA2012-32032, AYA2016-75066-C2-$1/2$-P, CSD2009-00038 (ASTROMOL) under the Consolider-Ingenio Program, and acknowledge funding support from the European Research  Council ERC  Grant  610256:  NANOCOSMOS.  R.B. and M.T. acknowledge support from grant AYA2016-79006-P.
EM acknowledges support from the Brazilian agency FAPESP, projects $2014/22095-6$ and
$2015/22254-0$.

\appendix
\section{Examples of chemical differentiation in the ASAI source sample}

\begin{figure*}
\center{
\caption{Comparison of CH$_2$CN (left) and CH$_3$CN (right) emission line profiles in the 3mm band. The frequencies of the transitions are marked with arrows.}
\includegraphics[width=\columnwidth]{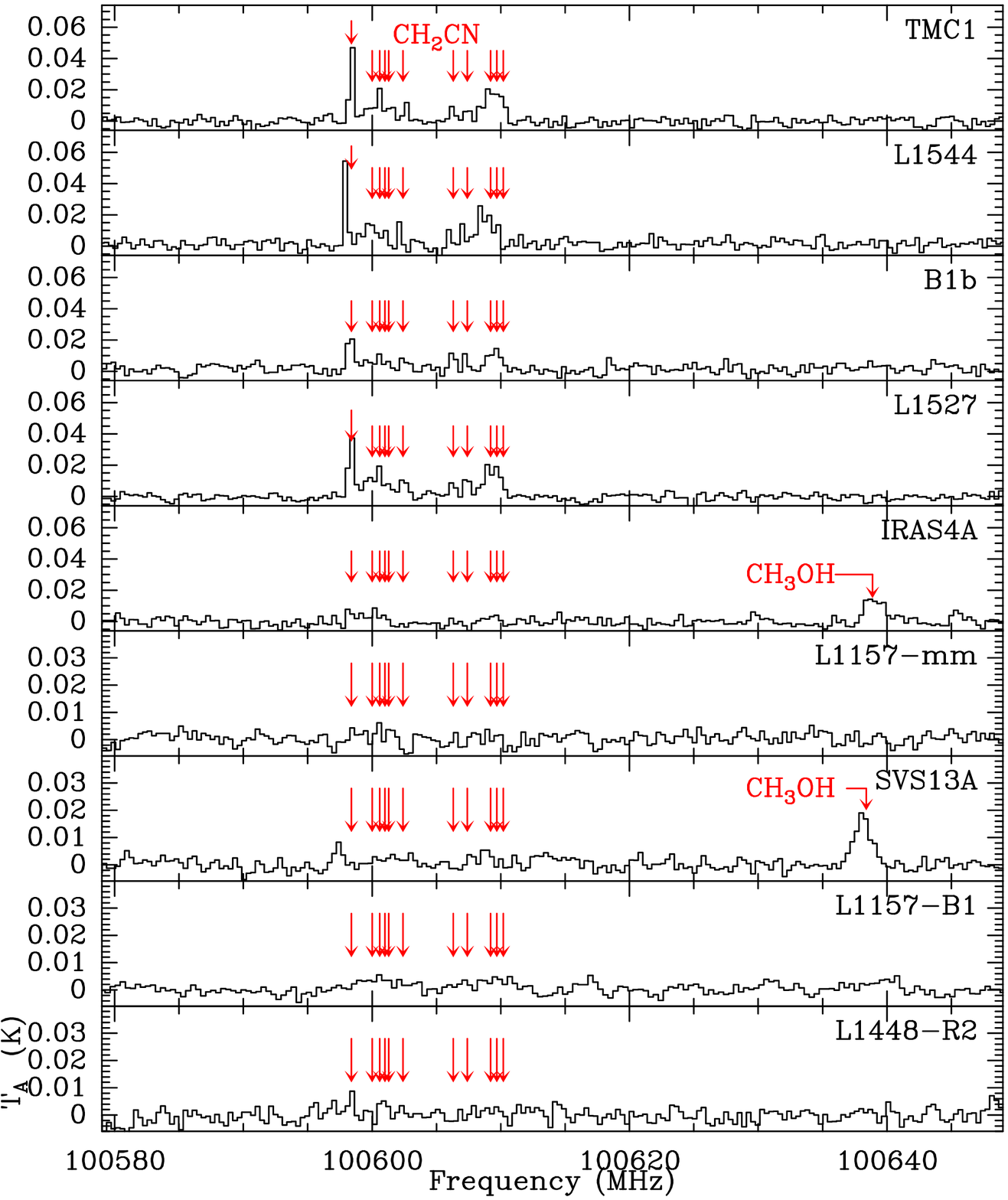}
\includegraphics[width=\columnwidth]{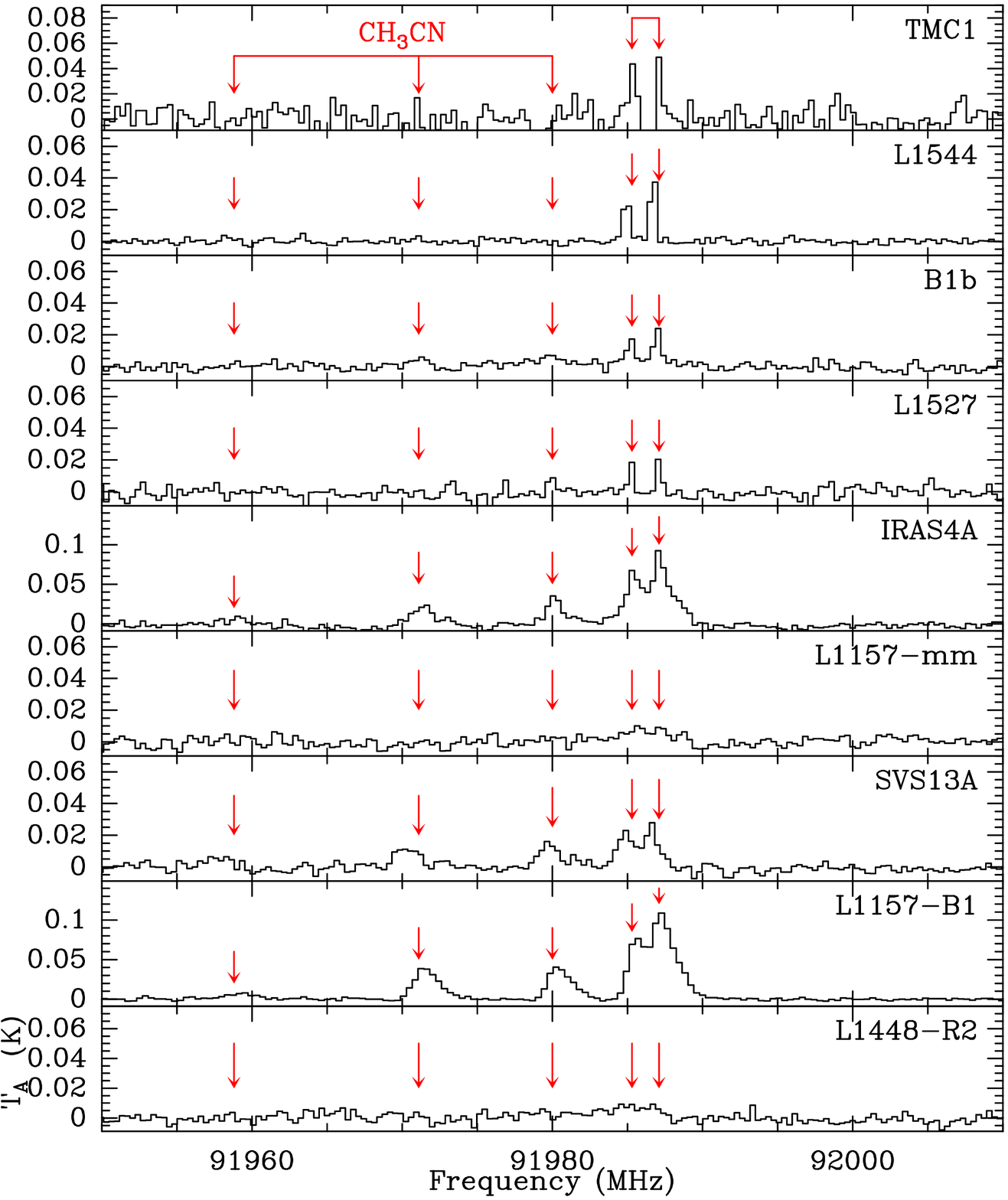}
}
\end{figure*}

\begin{figure}
\center{
\caption{Comparison of COM emission between 90100 and 90255 MHz.}
\includegraphics[width=\columnwidth]{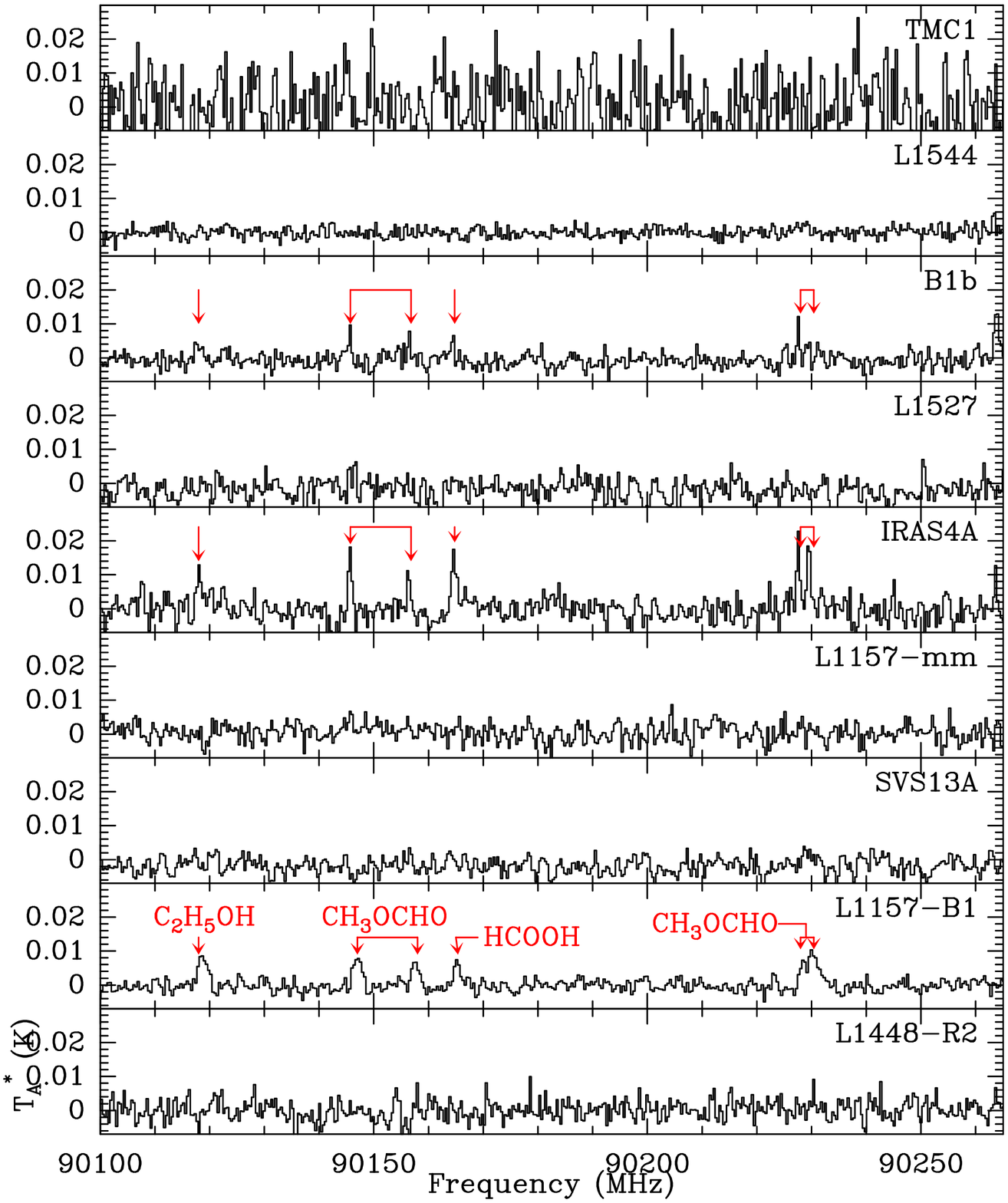}

}
\end{figure}

\end{document}